\documentclass[10pt]{article}

\usepackage[scaled]{helvet}

\usepackage{graphicx,natbib} 
\usepackage{enumerate}

\usepackage{url} 
\newcommand{\blind}{0}

\usepackage{tikz}
\usetikzlibrary{shapes.geometric, arrows}
\usepackage[doublespacing]{setspace} 
\usepackage{lineno} 
\usepackage{authblk} 
\usepackage{amsmath} 
\usepackage{bm} 
\usepackage{amssymb} 
\usepackage[inline]{enumitem} 
\usepackage{float}
\usepackage[labelsep=period,font=scriptsize]{caption} 
\usepackage[utf8]{inputenc} 
\usepackage[T1]{fontenc} 
\usepackage[title]{appendix}
\usepackage{multirow}
\usepackage{booktabs}
\usepackage{array}
\usepackage{booktabs,subcaption,amsfonts,dcolumn}
\usepackage{soul}
\newcolumntype{M}[1]{>{\centering\arraybackslash}m{#1}}
\usepackage{changes}
\newcommand{\stkout}[1]{\ifmmode\text{\sout{\ensuremath{#1}}}\else\sout{#1}\fi}
\setdeletedmarkup{\stkout{#1}}

\DeclareMathOperator{\E}{E}

\DeclareMathOperator*{\argmin}{argmin}  
 
\DeclareMathOperator*{\arginf}{arginf} 
\DeclareMathOperator{\diag}{diag}

\tikzstyle{process} = [rectangle, minimum width=3cm, minimum height=1cm, text centered, draw=black ]
\tikzstyle{arrow} = [thick,->,>=stealth]
\begin{document}

\def\spacingset#1{\renewcommand{\baselinestretch}%
{#1}\small\normalsize} \spacingset{1}


\if0\blind
{
  \title{\bf Statistical Process Monitoring based on\\ Functional Data Analysis}

\author[1]{Fabio Centofanti\thanks{Corresponding author. e-mail: \texttt{fabio.centofanti@kuleuven.be}}}

\affil[1]{Section of Statistics and Data Science, Department of Mathematics, KU Leuven, Belgium}

\setcounter{Maxaffil}{0}
\renewcommand\Affilfont{\itshape\small}
\date{}
\maketitle

} \fi

\if1\blind
{
  \begin{center}
    {\LARGE\bf An Adaptive Multivariate Functional Control Chart}
\end{center}
} \fi

\normalsize

\begin{abstract}
In modern industrial settings, advanced acquisition systems allow for the collection of data in the form of profiles, that is, as functional relationships linking responses to explanatory variables.
In this context, statistical process monitoring (SPM) aims to assess the stability of profiles over time in order to detect unexpected behavior.
This review focuses on SPM methods that model profiles as functional data, i.e., smooth functions defined over a continuous domain, and apply functional data analysis (FDA) tools to address limitations of traditional monitoring techniques.
 A reference framework for monitoring multivariate functional data is first presented. This review then offers a focused survey of several recent FDA-based profile monitoring methods that extend this framework to address common challenges encountered in real-world applications. These include approaches that integrate additional functional covariates to enhance detection power,  a robust method designed to accommodate outlying observations, a real-time monitoring technique for partially observed profiles, and two adaptive strategies that target the characteristics of the out-of-control distribution. 
These methods are all implemented in the \textsf{R} package \textsf{funcharts}, available on CRAN.
Finally, a review of additional existing FDA-based profile monitoring methods is also presented, along with suggestions for future research.
\end{abstract}
\noindent%
\vfill

\newpage

\spacingset{1.45}
\section{Introduction}
\label{sec_intro}

Nowadays, the development of modern acquisition systems enables the collection of data that are typically characterized by significant complexity and high dimensionality.
A particularly relevant case arises when data are recorded as profiles, that is, as functional relationships linking responses to explanatory variables.”
In this setting,  profile monitoring \citep{noorossana2011statistical} refers to the application of
 statistical process monitoring (SPM) methods \citep{montgomery2007introduction,qiu2013introduction}
to one or more quality characteristics of interest that are observed as profiles.
The aim of statistical process monitoring (SPM) is to assess the stability of a process. This is typically achieved through a control chart, which displays one or more statistics plotted over time, together with control limits that represent the expected range of variation due to common causes. Observations falling outside these limits indicate the possible presence of special causes of variation, in which case the process is deemed out-of-control (OC). Otherwise, the process is considered to be in control (IC).

 Because of its applicative relevance, profile monitoring has been extensively studied in recent years, and several methods have been developed. Traditional approaches treat each observation within a profile as an output for conducting regression. The resulting regression coefficients or residuals are then used to construct monitoring schemes \citep{woodall2004using,noorossana2011statistical}. 
 Some relevant works include the methods of \cite{mahmoud2004phase,wang2005using,zou2006control,williams2007statistical,zou2007monitoring, zou2008monitoring, qiu2010control,zhou2022phase,wang2022outlier}, which are based on multivariate control charts applied to the linear and nonlinear regression coefficients. \cite{qiu2010nonparametric,rajabi2017phase} presented methods
based on mixed-effect models, while  \cite{zou2012lasso} proposed a new methodology for monitoring general multivariate linear profiles via a LASSO-based multivariate SPM technique.  Comprehensive reviews of these approaches are provided in \cite{maleki2018overview} and \cite{liu2024comprehensive}.

While regression-based methods can be effective in many applications, they encounter significant limitations in more complex settings where profile data are characterized by nonstationary patterns, irregularly sampled observations, and strong within-profile and between-profile correlations. These issues can undermine the assumptions underlying such methods, making them inadequate for capturing the full complexity of the data.
These limitations can be overcome by modelling the observations as functional data, i.e.,  smooth functions defined over continuous domains.
Functional Data Analysis (FDA) offers a comprehensive statistical framework for modelling and analysing such data structures \citep{ramsay2005functional,ferraty2006nonparametric,horvath2012inference,kokoszka2017introduction}.

\begin{figure}[h]
    \centering
    \includegraphics[width=\textwidth]{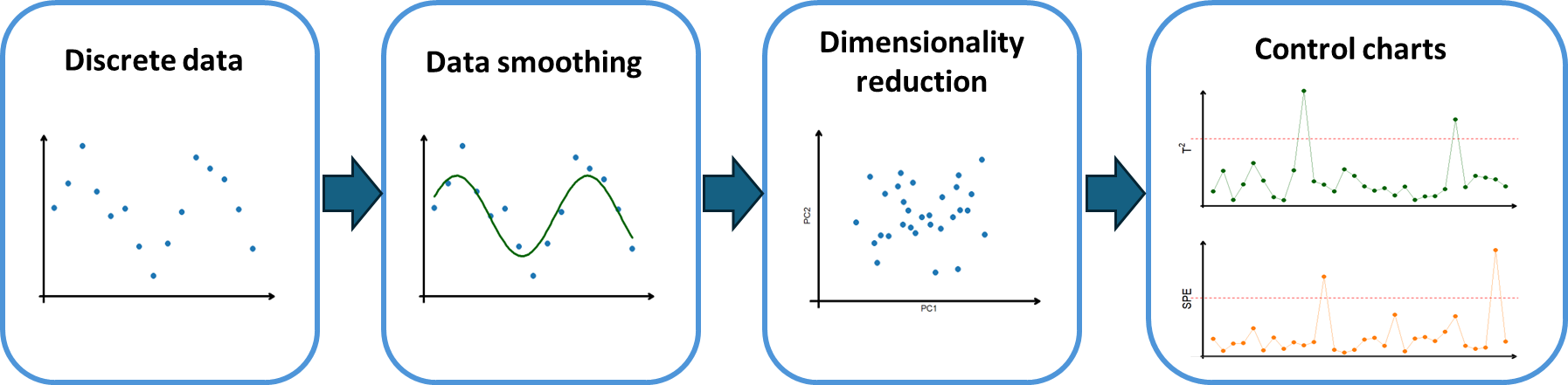}
    \caption{The reference framework for monitoring 
functional data.}
    \label{fig:structure}
\end{figure}

FDA is now a well-established branch of statistics, yet its integration into profile monitoring is comparatively recent and still limited. Although the past decade has seen a rapid growth of FDA‑based monitoring methods, their adoption in practice remains far behind that of traditional SPM techniques.
One possible reason is the fragmented literature, with methods scattered across disparate publications and rarely presented within a single coherent framework.
This review seeks to bridge this gap by synthesizing existing methods for monitoring multivariate functional data into a unifying reference framework. As detailed in Section \ref{sec_generalstrategy}, this framework is structured around three main components: data smoothing (Section \ref{sec_smoothing}), dimensionality reduction through multivariate functional principal components (Section \ref{sec_MFPCA}), and monitoring and diagnostic procedures (Section \ref{sec_monitoring}). 
The overall monitoring pipeline is illustrated in Figure \ref{fig:structure}.
The emphasis is on Phase II monitoring. In this setting, a reference dataset of observations, which is assumed to represent IC conditions, is first used in Phase I to estimate unknown parameters. In Phase II, incoming observations are monitored in real time.
In Section \ref{sec_exam}, an illustrative example is presented in the monitoring of ship operating conditions.
Moreover, to facilitate the wider use of FDA‑based profile monitoring techniques, a focused survey of several recently developed methods is also provided.
These methods adapt and extend the reference framework for monitoring functional data to address specific challenges commonly encountered in real-world applications. In particular, several approaches are discussed that incorporate additional functional covariates to improve monitoring performance (Section \ref{sec_cov}), a robust method designed to handle outlying observations (Section \ref{sec_out}), a real-time monitoring technique applicable when the functional quality characteristic is only partially observed (Section \ref{sec_FRTM}), and two strategies that adapt to the unknown distributions of the OC data (Section \ref{sec_adaptive}).
The lack of accessible software tools is another key factor limiting the adoption of FDA-based profile monitoring approaches in practice.
However, these methods can be directly applied in practice, as they are all readily implemented in the \textsf{R} package \textsf{funcharts}, available on CRAN \citep{capezza2025funcharts}. 
Section \ref{sec_additional} reviews additional FDA-based profile monitoring methodologies, aiming to provide a comprehensive overview of the current literature.
Finally, Section \ref{sec_conclusion} presents the concluding remarks and suggestions for future research.

\section{A Reference Framework for Monitoring Multivariate Functional Data}
\label{sec_generalstrategy}
In this section, the reference framework for monitoring multivariate functional data is reviewed.
This framework integrates several specific contributions proposed in \cite{capezza2021functional, centofanti2021functional, centofanti2025adaptive, capezza2023funcharts, centofanti2025real, capezza2025adaptive}. 
Section \ref{sec_smoothing} presents a data smoothing approach via a roughness penalty to obtain multivariate functional data from noisy discrete measurements. Section \ref{sec_MFPCA} describes the multivariate functional principal component analysis (MFPCA) that allows reducing the infinite-dimensional nature of the data.
These elements are used in Section \ref{sec_monitoring} to construct monitoring and diagnostic procedures based on Hotelling $T^2$ and $SPE$ statistics.
In Section \ref{sec_exam}, this framework is applied for monitoring the operating conditions of a Ro-Pax ship.

\subsection{Data Smoothing}
\label{sec_smoothing}
The $p$-dimensional multivariate functional quality characteristic   $\bm{X}=\left(X_1,\dots, X_p\right)^{T}$ is assumed to be a random vector with realizations in the space of $p$-dimensional vectors of $L^2(\mathcal{T})$ functions.
The $L^2(\mathcal{T})$ space is the space of square-integrable functions defined over the compact domain $\mathcal T \subset \mathbb{R}$.
Data are usually collected discretely, which means that any component of the multivariate functional quality characteristic $\bm{X}$ is observed through the values $\lbrace Y_{k}\left(t_{kj}\right), t_{kj}\in\mathcal{T}, j=1,\dots,n_{k} \rbrace_{k=1,\dots,p}$.
In case of no measurement error, multivariate functional data are obtained by connecting the set of points $ Y_{k}\left(t_{kj}\right)$, $j=1,\dots,n_{k}$, for each $k$. However, this does not represent the ordinary situation where observational error superimposes on the underlying signal $\bm{X}$, that is
\begin{equation*}
Y_{k}\left(t_{kj}\right)=X_{k}\left(t_{kj}\right)+\varepsilon_{kj}, \quad t_{kj}\in\mathcal{T}, j=1,\dots,n_{k}, k=1,\dots,p,
\end{equation*}
where $\bm{\varepsilon}_{k}=\left(\varepsilon_{k1},\dots,\varepsilon_{kn_{k}}\right)^{T}$ are zero mean random errors vectors.
To recover the the underlying signal,  a common approach consists of representing each $X_{k}$   
as a linear combination of $K_k$ known basis functions $\phi_{k1},\dots,\phi_{kK_k}$ as follows
\begin{equation}
\label{eq_fk}
	X_{k}\left(t\right)=\sum_{m=1}^{K_k}c_{km}\phi_{km}\left(t\right),\quad  t\in\mathcal{T}, k=1,\dots,p.
\end{equation}
Then,  the unknown coefficient vectors $\bm{c}_{k}=\left(c_{k1},\dots,c_{kK_k}\right)^{T}$ are estimated as
\begin{multline}
	\label{eq_smootspline2}
\argmin_{\bm{c}_1\in\mathbb{R}^{K_1},\dots,\bm{c}_p\in\mathbb{R}^{K_p}}\sum_{k_1=1}^p\sum_{k_2=1}^p\left(\bm{y}_{k_1}-\bm{c}_{k_1}^T\bm{T}_{k_1}\right)^{T}\bm{D}_{k_1k_2}\left(\bm{y}_{k_2}-\bm{c}_{k_2}^T\bm{T}_{k_2}\right)+\sum_{k=1}^p\lambda_k\bm{c}_{k}^T\bm{R}_{k}^{(m)}\bm{c}_{k},
\end{multline}
where $\bm{y}_{k}=\left(Y_{k}(t_{k1}),\dots, Y_{k}(t_{kn_{k}})\right)^{T}$, $\bm{T}_{k} $ contains the  values of the $K_k$ basis functions evaluated at $t_{k1}\dots,t_{kn_{k}}$, and   $\bm{R}^{(m)}_k $ is a matrix whose entry $i,j$ is $\int_{\mathcal{T}} \phi^{\left(m\right)}_{ki}(t)\phi^{\left(m\right)}_{kj}(t)dt$, where  $\phi^{\left(m\right)}_{ki}$ is the $m$-derivative of $\phi_{ki}$. The positive definite symmetric matrices $\bm{D}_{k_1k_2}$ allow for unequal weighting of the squared residuals. The smoothing parameters $\lambda_k$ control the trade-off between goodness-of-fit and the smoothness of the final estimates, where smoothness is quantified by the quadratic penalty term in the second component on the right-hand side of \eqref{eq_smootspline2}.
As basis functions $\phi_{k1},\dots,\phi_{kK_k}$, the B-spline basis functions are the most common choice because of good computational properties and great flexibility \citep{ramsay2005functional}. The penalty on the right-hand side of \eqref{eq_smootspline2}  is computed by setting  $m=2$, i.e.,  by penalizing the function's second derivatives.
The values  $K_k$ in Equation \eqref{eq_fk}  are not crucial \citep{cardot2003spline}, and $\bm{D}_{k_1k_2}$ are usually assumed to be the identity matrices.
The smoothing parameters $\lambda_k$ in \eqref{eq_smootspline2} are selected by minimizing the Generalized Cross-Validation (GCV) criterion \citep{ramsay2005functional}.

\subsection{Multivariate Functional Principal Component Analysis}
\label{sec_MFPCA}
 The MFPCA is used to reduce the infinite dimensionality of $\bm{X}$. 
 Differences in variability and unit of measurements among $X_1,\dots, X_p$ are taken into account by considering the standardized  $\bm{Z}=\left(Z_1,\dots,Z_p\right)^T$, with $Z_k(t)=(v_k(t))^{-1/2}(X_k(t)-\mu_k(t))$,  where $\mu_k$ and $v_k$ are the mean  are variance functions of each component. 
From the multivariate Karhunen-Lo\`{e}ve's theorem \citep{happ2018multivariate}, it follows that
\begin{equation*}
	\bm{Z}(t)=\sum_{l=1}^{\infty} \xi_l\bm{\psi}_l(t),\quad t\in\mathcal{T},
\end{equation*}
where $\bm{\psi}_l=\left(\psi_{l1},\dots,\psi_{lp}\right)^T$, and $\xi_l=\sum_{k=1}^{p}\int_{\mathcal{T}}  Z_k(t)\psi_{lk}(t)dt$ are  called principal component scores, such that  $\E\left( \xi_l\right)=0$ and $\E\left(\xi_l \xi_m\right)=\eta_{l}\delta_{lm}$, being $\delta_{lm}$ the Kronecker delta.
The vectors of function  $\bm{\psi}_l $, with $\sum_{k=1}^{p}\int_{\mathcal{T}} \psi_{lk}(t)\psi_{mk}(t)dt=\delta_{lm}$, are referred to as principal components and are the eigenfunctions of the covariance $\bm{C}$  of $\bm{Z}$ corresponding to the eigenvalues $\eta_1\geq\eta_2\geq \dots\geq 0$, that is 
\begin{equation*}
\int_{\mathcal{T}}\bm{C}(s,t)\bm{\psi}_l(s)ds=
	\eta_l\bm{\psi}_l(t),\quad t\in\mathcal{T},
\end{equation*}
where $\bm{C}=\lbrace C_{k_1k_2}\rbrace_{1\leq k_1,k_2 \leq p}$, $C_{k_1k_2}(s,t)=\E(Z_{k_1}(s),Z_{k_2}(t))$, $s,t\in \mathcal{T}$.
To estimate the eigenfunctions and eigenvalues of $\bm{C}$, 
each  component of the eigenfunction $\bm \psi_l$  is represented as a linear combination of the $K_k$  basis functions $\phi_{k1},\dots,\phi_{kK_k}$ used in \eqref{eq_fk} 
to obtain the  estimates $\bm{X}$, that is
\begin{equation}
	\label{eq_appcov}
	\psi_{lk}(t)= \sum_{m=1}^{K_k} b_{lkm}\phi_{km}(t), \quad  t\in\mathcal{T}, k=1,\dots,p,
\end{equation}
where $\bm{b}_{lk}=\left(b_{lk1},\dots,b_{lkK_k}\right)^T$ are coefficient vectors of  the eigenfunction. 
Then, the eigenfunctions and eigenvalues of the covariance $\bm{C}$ are estimated by applying the classical multivariate principal component analysis to $\bm{W}^{1/2}\widetilde{\bm{c}}$, where $\widetilde{\bm{c}}$ is defined as the concatenation of the coefficient vectors associated with $\bm{Z}$, and $\bm{W}$ is a block-diagonal matrix with diagonal blocks $\bm{W}_k$, $k=1,\dots,p$, whose entries are $w_{k_1 k_2} = \int_{\mathcal{T}} \phi_{k_1}(t)\phi_{k_2}(t)dt$, $k_1,k_2=1,\dots, K_k$. 
A low-dimensional vector representation of the original data is obtained by retaining  the $L$ leading  eigenfunctions, which reflect the most important features of $\bm{Z}$.
This means that  $\bm{Z}$ is approximated as $\bm{Z}^L=\left(Z^L_1,\dots,Z^L_p\right)^T$ through the following truncated principal component decomposition
\begin{equation}
	\label{eq_appx}
	\bm{Z}^L(t)=\sum_{l=1}^{L} \xi_{l}\bm{\psi}_{l}(t), \quad t\in\mathcal{T}.
\end{equation}
The parameter $L$ is chosen such that the principal components retained explain at least a given percentage of the total variability, which is usually in the range 70-90$\%$ \citep{paynabar2016change,ren2019phase,centofanti2025real,capezza2024robust}, but more sophisticated methods can be used as well \citep{capezza2020control}.
\subsection{Monitoring and Diagnosis}
\label{sec_monitoring}
The information provided by   MFPCA is used to monitor $\bm{X}$ over time using the $T^2$ and the $SPE$ statistics, that are defined as
  \begin{equation}
      \label{eq_T2}T^2=\sum_{l=1}^{L}\frac{\xi_{l}^2}{\eta_l}, \quad SPE=\sum_{k=1}^{p}\int_{\mathcal{T}}\left(Z_k\left(t\right)-Z^L_k\left(t\right)\right)^2dt.
  \end{equation}
  Here, the $T^2$ statistic represents the squared Mahalanobis distance of the projection of $\bm{Z}$ onto the space spanned by the first $L$ principal components ${\bm{\psi}_l}$, while the $SPE$ quantifies the reconstruction error, i.e., the distance between $\bm{Z}$ and its truncated projection.
An OC signal is triggered if either $T^2 > CL^{T^2}$ or $SPE > CL^{SPE}$, where $CL^{T^2}$ and $CL^{SPE}$ are the control chart limits.

In Phase I, an IC dataset of discrete realizations of $\bm{X}$ is available. 
First, data smoothing is applied to obtain the multivariate functional data. 
Then, to reduce the undesirable effects caused by uncertainty in the estimation of the MFPCA model and to enhance monitoring performance \citep{ramaker2004effect,kruger2012statistical}, 
the dataset is partitioned into a training set and a tuning set, used for estimating the model parameters and control limits, respectively.
The model parameters include the variance functions $v_k$, the mean functions $\mu_k$, the principal components $\bm{\psi}_l$, and the corresponding eigenvalues $\eta_l$.
The control limits $CL^{T^2}$ and $CL^{SPE}$ are estimated as the $(1 - \alpha)$-quantiles of the empirical distribution of the two statistics computed on the tuning set.
The parameter $\alpha$ is chosen by using the Bonferroni correction $\alpha=\alpha^* / 2$, where $\alpha^*$ is the overall type I error probability.

After a signal, it is often critical to identify the components responsible for the OC condition. This is done by looking at the contributions of each functional variable to the $T^2$ and $SPE$ statistics, which are defined as
\begin{multline}
\label{eq_cont}
C^{T^2}_{k} = \sum_{l=1}^L\frac{\xi_{l}}{\eta_l} \int_{\mathcal T} Z_{k}(t) \psi_{lk}(t) dt, \quad C^{SPE}_{k} = \int_{\mathcal T} (Z_{k} (t) -  Z_{k}^L (t))^2 dt, \quad k=1,\dots, p.
\end{multline}
Contributions that exceed empirical quantiles, which are estimated in Phase I, indicate functional variables exhibiting unexpected variability.

\subsection{An Illustrative Example: Monitoring Ship Operating Conditions}
\label{sec_exam}
\begin{figure}[h]
\centering \includegraphics[width=0.45\textwidth]{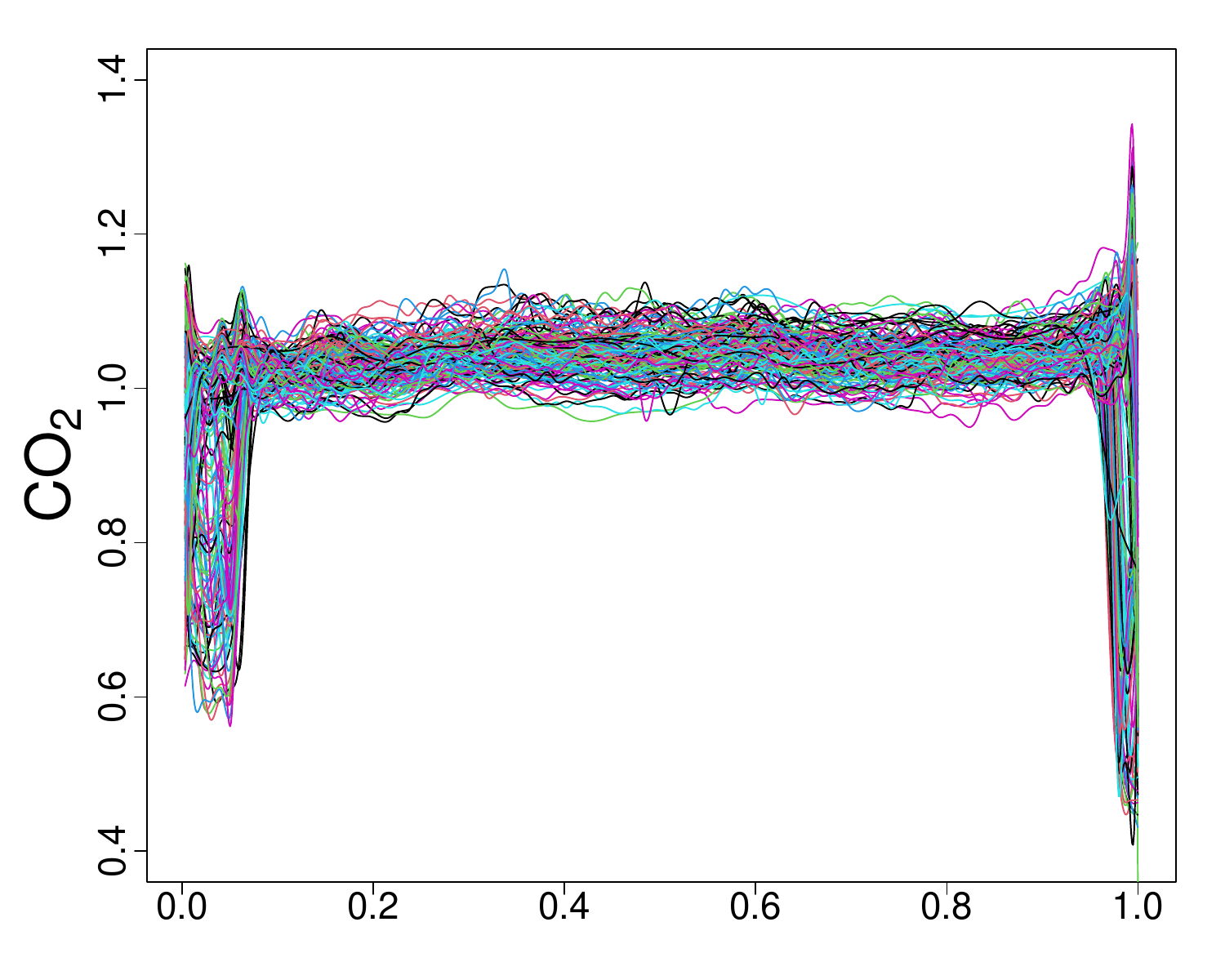}
\includegraphics[width=0.45\textwidth]{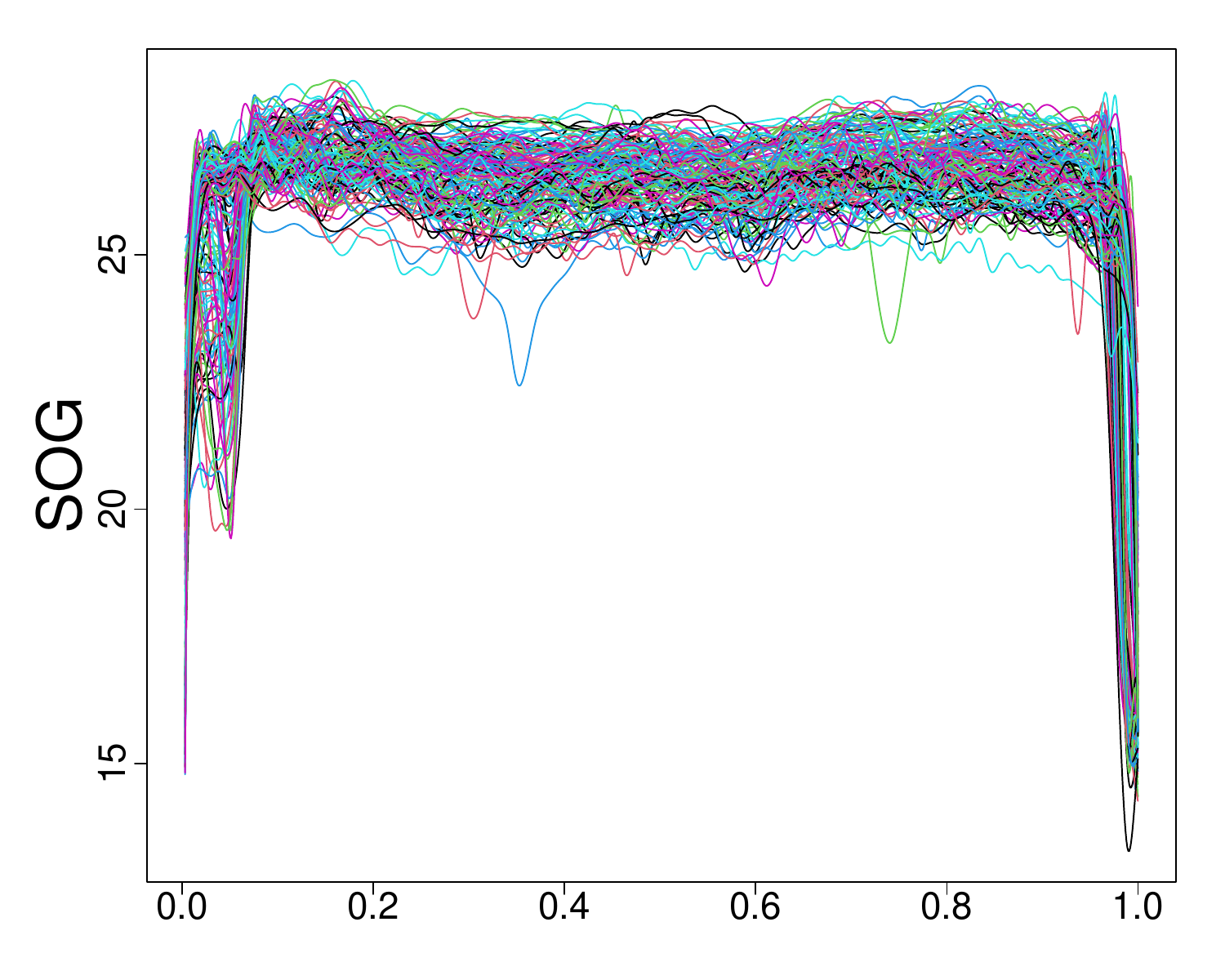}
\includegraphics[width=0.45\textwidth]{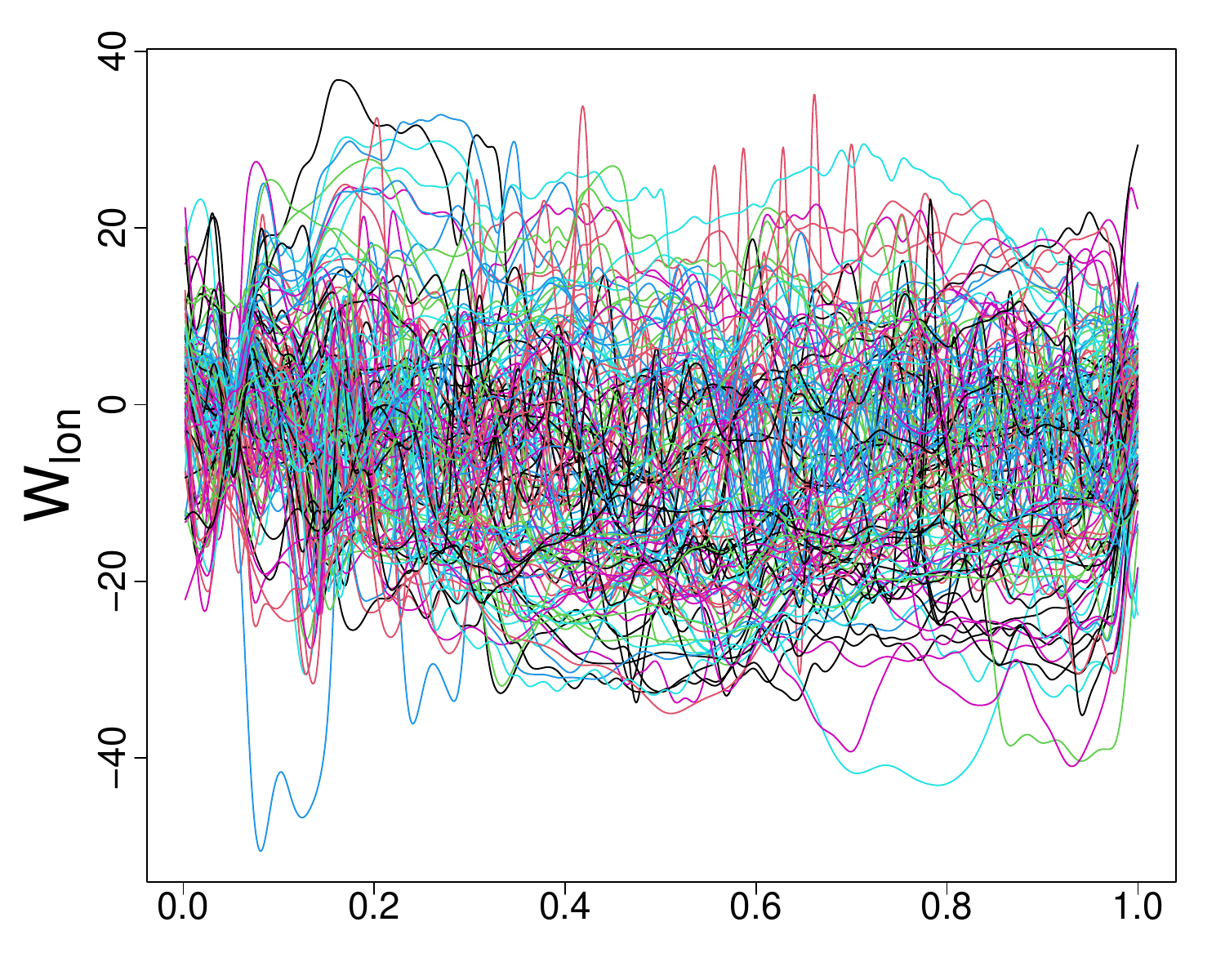}
\includegraphics[width=0.45\textwidth]{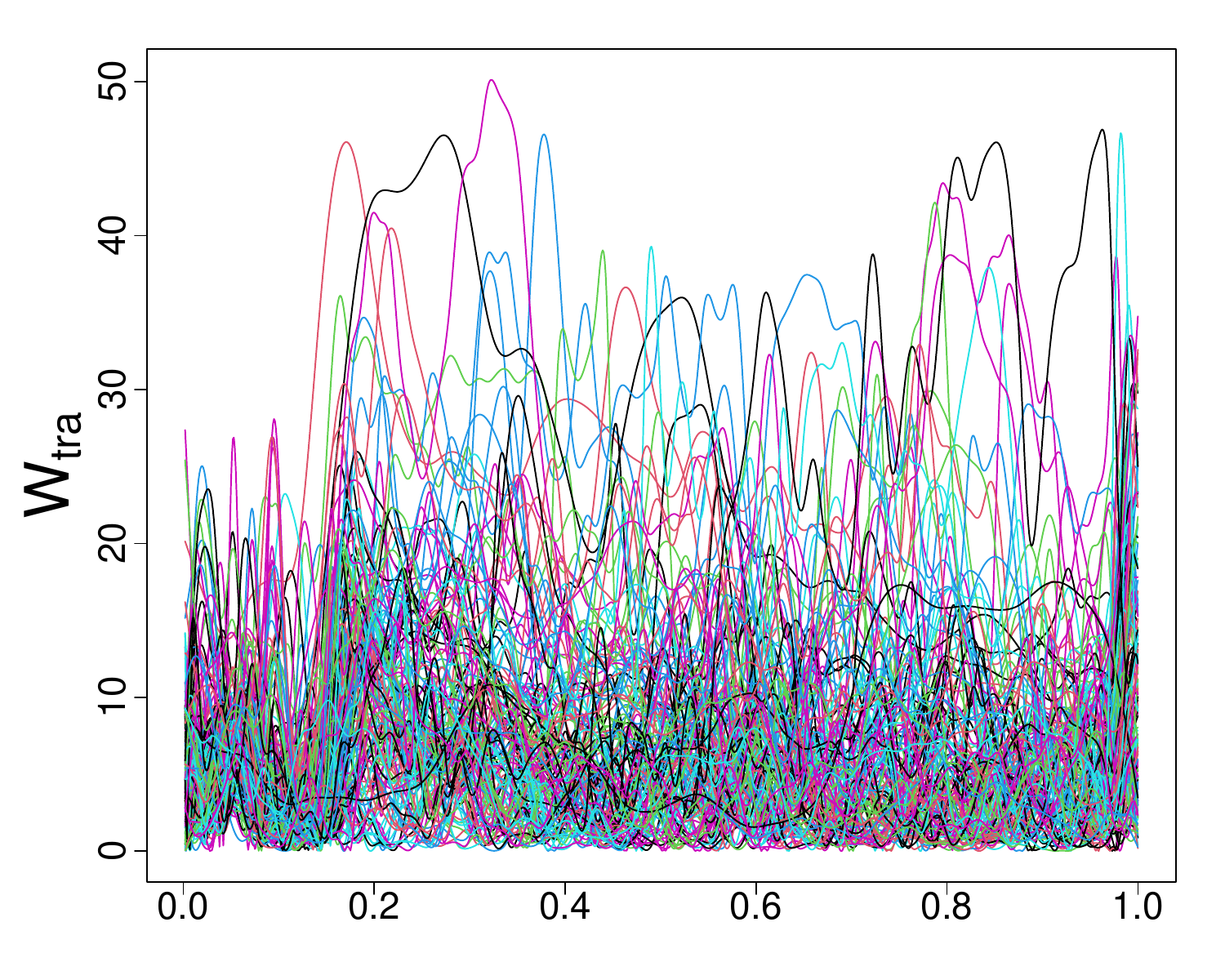}
\caption{\label{fig_data} Phase I data from the ship operating condition monitoring example. 
}
\end{figure}

In this section, the reference monitoring framework is applied to the SPM of operating conditions of a Ro-Pax ship, a vessel designed to carry both passengers and wheeled cargo such as cars and trucks. This topic has become 
extremely relevant in recent years, also in view of the dramatic climate change \citep{capezza2021functional,centofanti2021functional,capezza2022functional}.
The \texttt{ShipNavigation} dataset \citep{capezza2023funcharts} is considered, which includes, for each ship voyage, measurements of CO\textsubscript{2} emissions per nautical mile, speed over ground (SOG),  and the longitudinal (W\textsubscript{lon}) and transverse (W\textsubscript{tra}) wind components. The Phase I data, which are shown in Figure \ref{fig_data}, consists of measurements corresponding to 144 voyages, of which 104 are used for training and the remaining 40 for tuning. An additional set of 50 voyages along the same route is used in Phase II.

\begin{figure}[h]
\centering \includegraphics[width=0.9\textwidth]{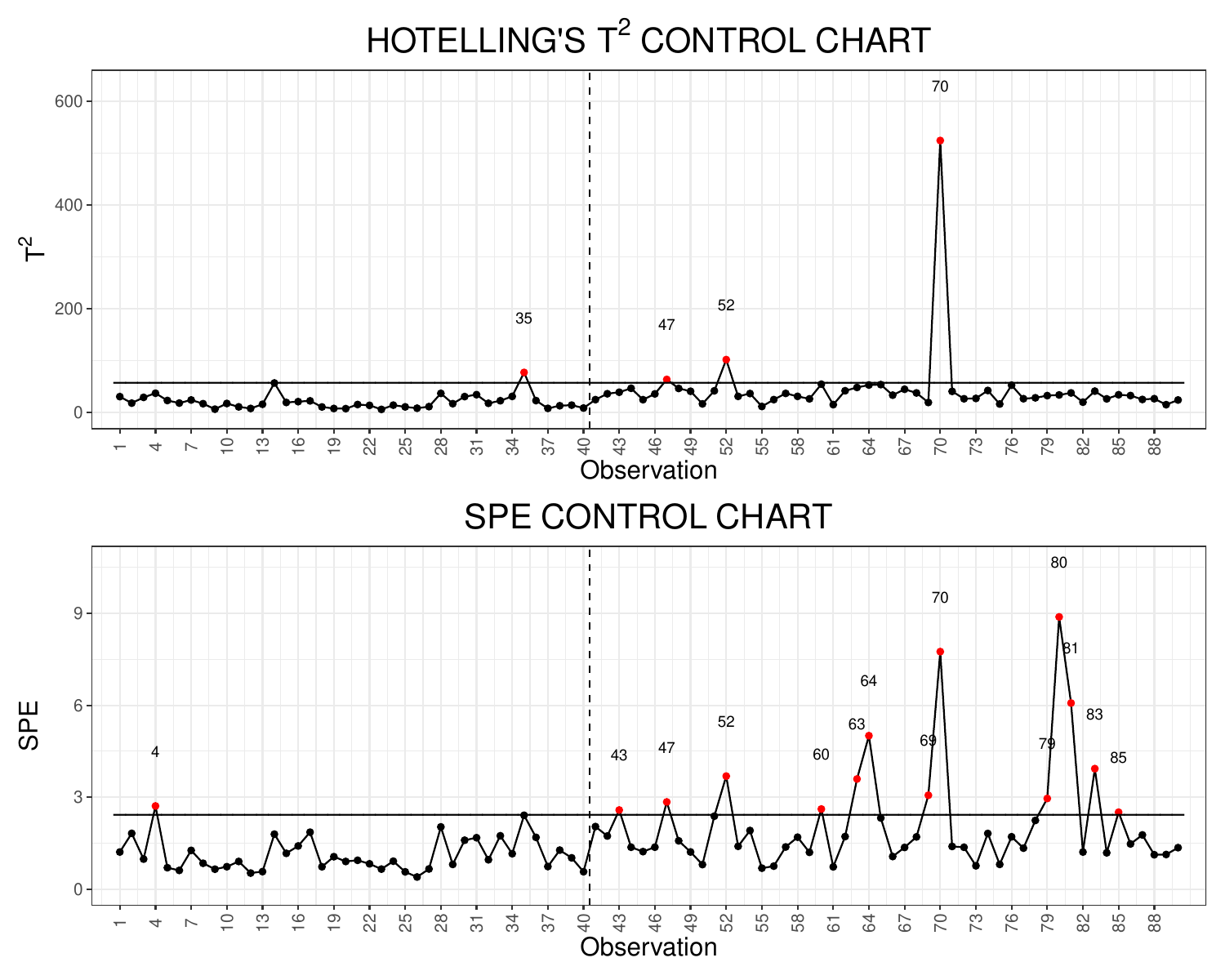} 
\caption{\label{fig:cc_real} 
Hotelling's $T^2$ and $SPE$ control charts for monitoring ship operating conditions. The vertical dashed line separates the tuning dataset from the Phase II dataset. Red points and associated observation indices indicate monitoring statistics that exceed the corresponding control limits.}
\end{figure}

\begin{figure}[h]
\centering \includegraphics[width=0.3\textwidth]{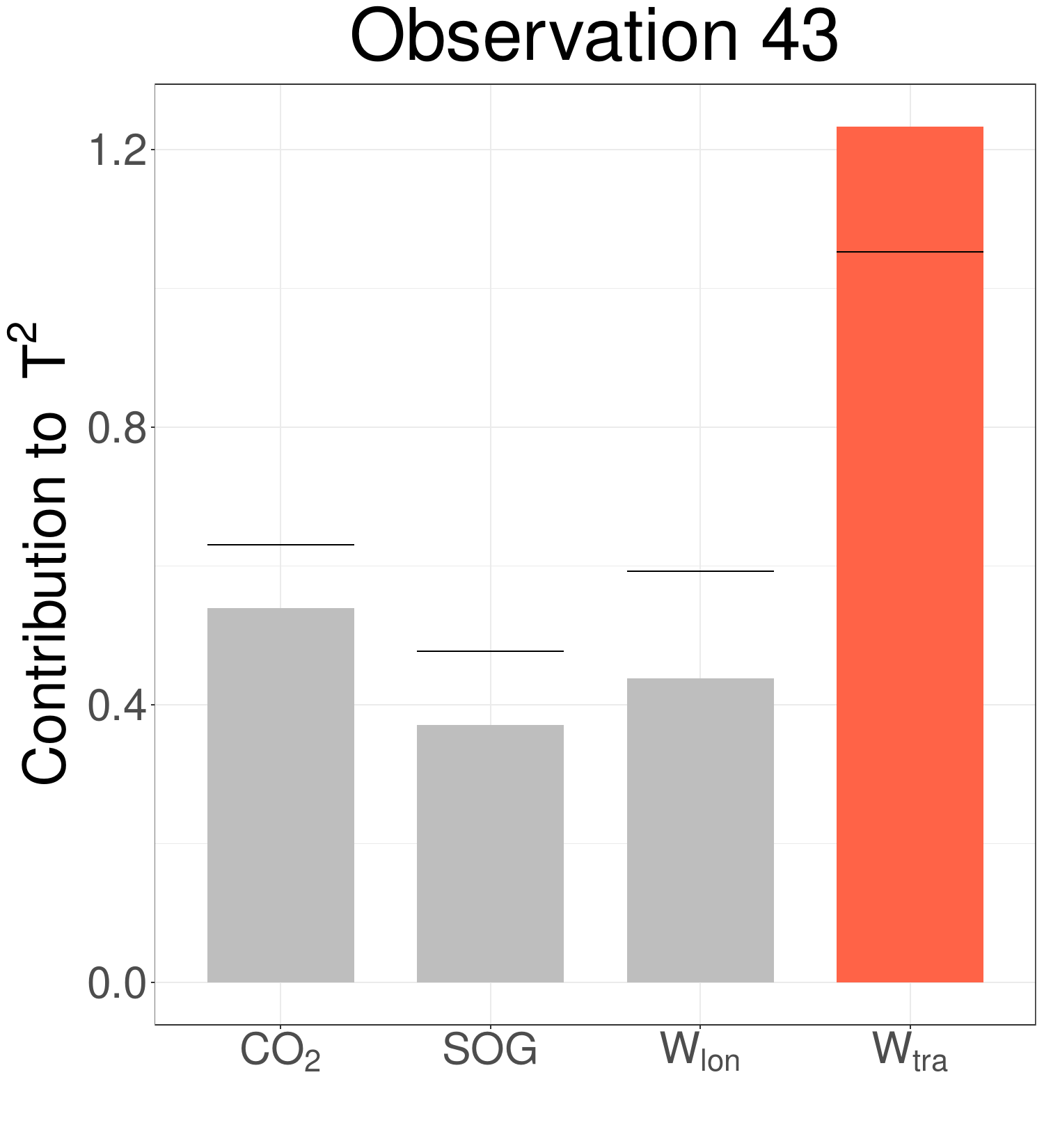} \includegraphics[width=0.3\textwidth]{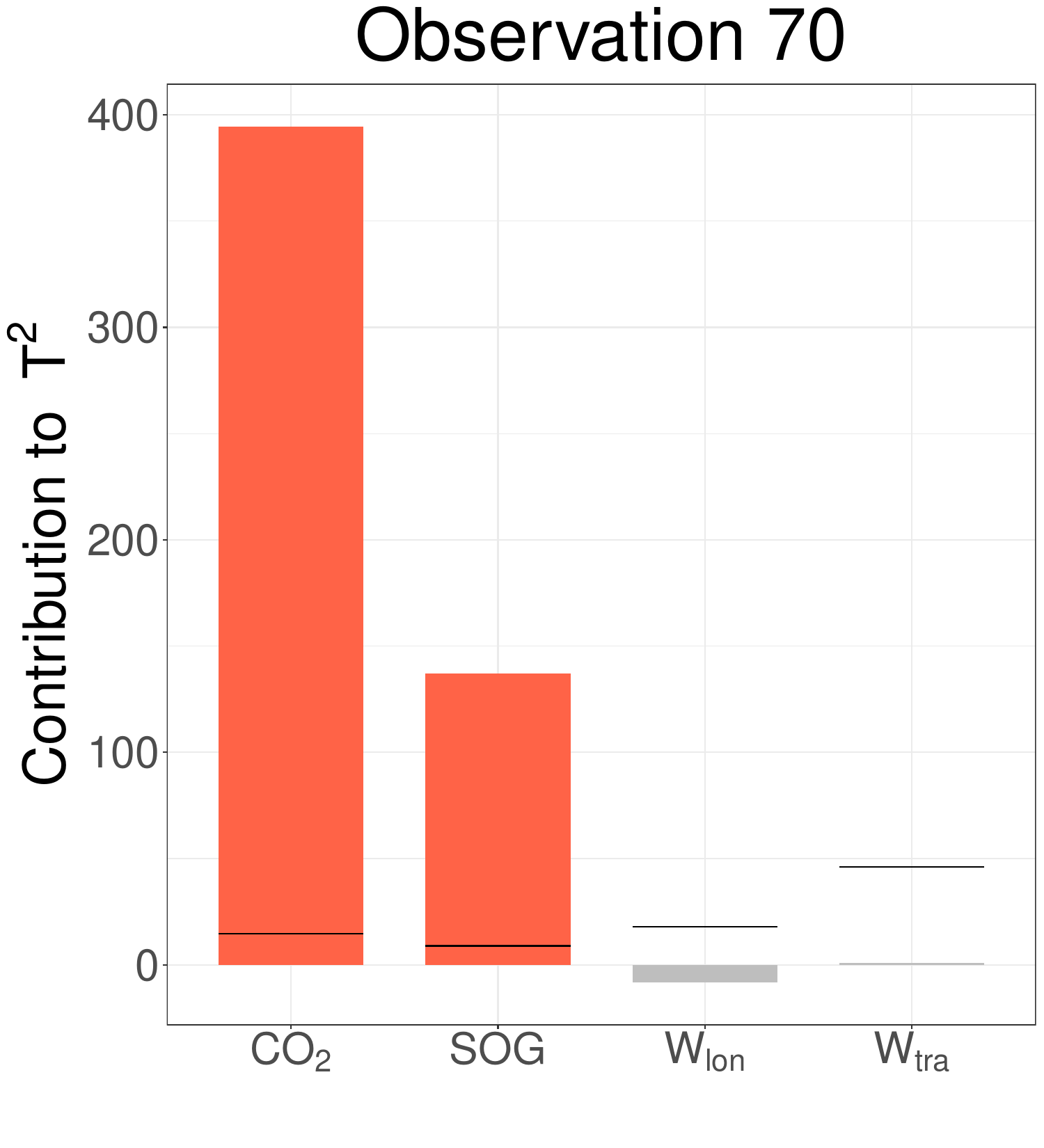} \includegraphics[width=0.3\textwidth]{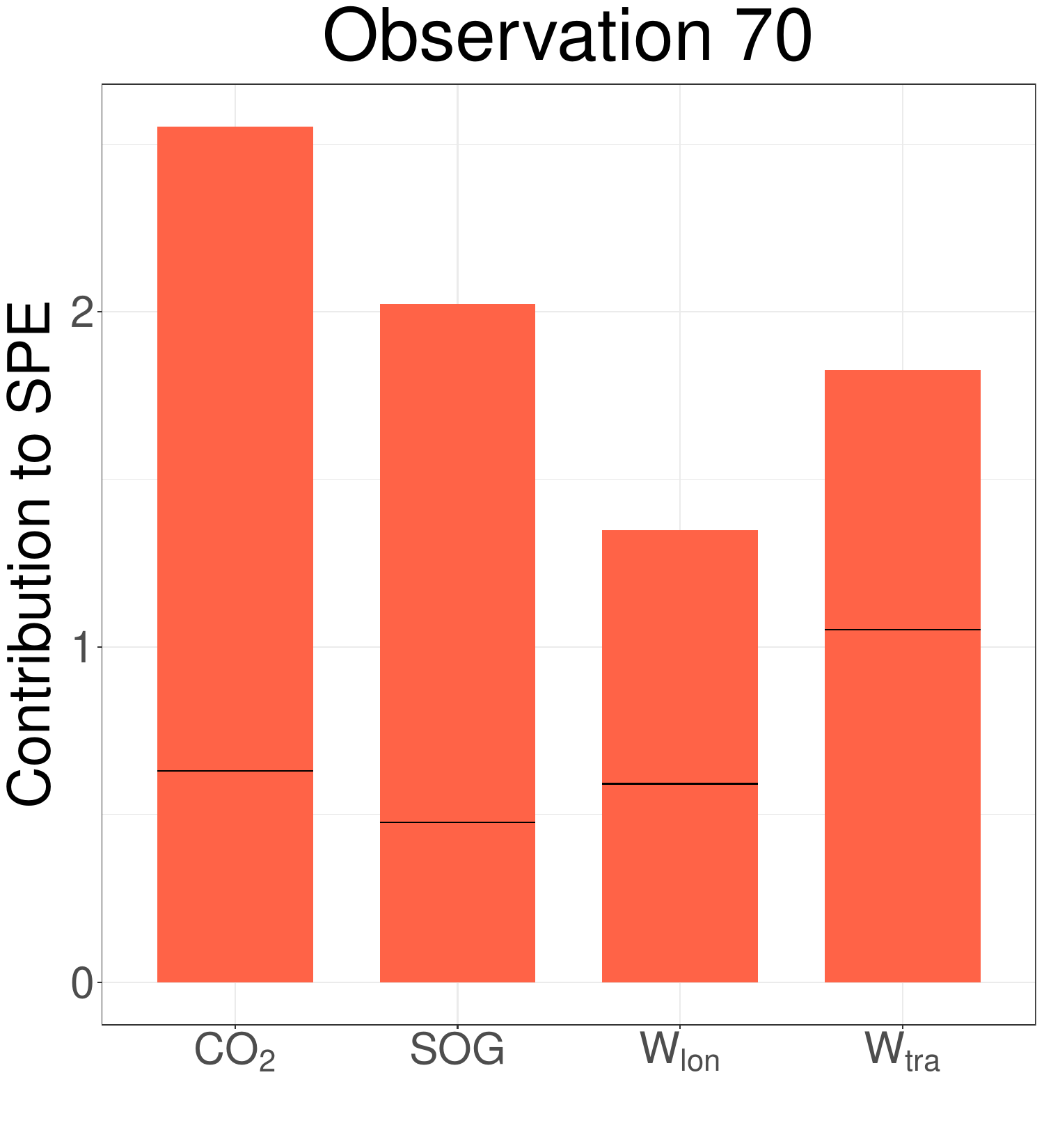} 
\caption{\label{fig_cont} 
Contributions to the Hotelling's $T^2$ and $SPE$ monitoring statistics corresponding to the OC observations 43 and 70. Contributions exceeding control limits, represented by the horizontal lines, are plotted in red. }
\end{figure}

Figure \ref{fig:cc_real} plots the monitoring statistics on the $T^2$ and $SPE$ control charts.  The monitoring statistics calculated on the tuning data set are plotted on the left-hand side of the vertical dashed line in grey. These results are obtained by applying the \textsf{R} package \textsf{funcharts},  available on CRAN \citep{capezza2025funcharts}.
Several voyages exhibit monitoring statistic values that exceed the corresponding control limits, indicating potential anomalies in the operating conditions.
Figure \ref{fig_cont} displays the signal contributions to the Hotelling's $T^2$ and $SPE$ monitoring statistics corresponding to the OC observations 43 and 70.
Observation 43 shows an anomalous value of $T^2$, likely caused by abnormal transverse wind components, whereas observation 70 presents anomalous values in both $T^2$ and $SPE$, driven by all the variables.
\section{Covariate-Adjusted Functional Data Monitoring}
\label{sec_cov}
In some applications, measurements of additional covariates are available alongside the quality characteristic. When the goal is to monitor the quality characteristic conditional on covariate values, the reference framework reviewed in Section \ref{sec_generalstrategy} is not applicable.
Indeed, if one of these covariates manifests itself with an extreme realization, the quality characteristic may wrongly be judged to be OC. Otherwise, there may be situations where the covariates are not extreme, and the quality characteristic may wrongly appear IC because the variance explained by the covariates is overlooked.

One proposed solution is the functional regression control chart (FRCC), introduced by \cite{centofanti2021functional}. 
It considers the case where the quality characteristic $Y$ is a univariate random function with values in $L^2(\mathcal{T})$, and the covariates are represented by a $p$-dimensional random vector $\bm{X}=\left(X_1,\dots,X_p\right)^T$ of functions in $L^2(\mathcal{S})$, defined over a compact domain $\mathcal{S} \subset \mathbb{R}$.
Without loss of generality, both $Y$ and $\bm{X}$ are centred and scaled as described in Section \ref{sec_MFPCA}.
The FRCC is based on the following three main steps, in which one should define ($i$) the model that links $Y$ and $\bm{X}$; ($ii$) the estimation method for the chosen model;  ($iii$)  the monitoring strategy of the functional residuals.
\cite{centofanti2021functional} consider a specific implementation of the FRCC framework in which step (i) adopts the function-on-function linear regression model 
\begin{equation}
\label{eq_fof}
Y(t) = \sum_{k=1}^{p} \int_{\mathcal{S}} X_k(s) , \beta_k(s, t) , ds + \varepsilon(t), \quad t \in \mathcal{T},
\end{equation}
where $\beta_1, \dots, \beta_p$ are bivariate square integrable functional coefficients defined on the domain $\mathcal{S} \times \mathcal{T}$, and $\varepsilon$ denotes a zero mean functional error term defined over  $\mathcal{T}$.
Then,  the step ($ii$) is performed by applying MFPCA to both the set of functional covariates and to the response function separately, as described in Section \ref{sec_MFPCA}, yielding respectively eigenfunctions $\bm \psi^X_{l} = (\psi^X_{l1}, \dots \psi^X_{lp})^T $ and $ \psi^Y_m$ with the corresponding eigenvalues $\eta^X_l$ and $\eta^Y_m$. 
The covariates $X_{k}$ and the response $Y$ can be then  approximated as
\begin{equation*}
 X^L_{k}(s) = \sum_{l=1}^L \xi^X_{l} \psi^X_{lk}(s), \quad  Y^M(t) = \sum_{m=1}^M\xi^Y_{m} \psi^Y_{m}(t),  \quad s \in \mathcal S, \quad t \in \mathcal T, \quad k=1,\dots, p,
\end{equation*}
where $\xi^Y_{m}=\int_{\mathcal T} Y(t) \psi_{m}^Y (t) dt$, $\xi^X_{l}=\sum_{k=1}^p \int_{\mathcal S} X_{k}(s) \psi_{lk}^X (s) ds$, and $L$ and $M$ denote the number of retained principal components to represent $X_{k}$ and  $Y$, respectively.
Similalry,  the functional coefficients $ \beta_1,\dots, \beta_p$ in  \eqref{eq_fof} are represented  through the following tensor product basis expansion
\begin{equation}
\label{eq_fof_lm}
\beta^{LM}_k(s, t) = \sum_{l=1}^L \sum_{m=1}^M b_{lm} \psi^X_{mk} (s) \psi^Y_l(t), \quad s \in \mathcal S, \quad t \in \mathcal T, \quad k = 1, \dots, p,
\end{equation}
where $b_{lm}$ are the basis coefficients.
Then, model \eqref{eq_fof} reduces to
\begin{equation}
\label{eq_fof_simplified}
\xi^Y_{m} = \sum_{l=1}^L \xi^X_{l} b_{lm} + \epsilon_{m}, \quad m=1,\dots,M, 
\end{equation}
with   $\epsilon_m=\int_{\mathcal{T}}\varepsilon\left(t\right)\psi^{Y}_m\left(t\right)dt$.
The basis coefficients $b_{lm}$ are obtained through least squares as $\hat b_{lm} = \E (\xi^Y_{m} \xi^X_{l}) / \E\left[ (\xi^X_{l})^2\right]$, which plugged into \eqref{eq_fof_lm} and \eqref{eq_fof} produce the estimate $\hat \beta_1, \dots, \hat \beta_p$ and the fitted value $\hat Y$ of  $Y$.
In the final step ($iii$), the FRCC computes the functional residual as
\begin{equation}
\label{eq_fof_res}
e(t) = Y(t) - \hat{Y}(t), \quad t \in \mathcal{T}.
\end{equation} 
\cite{centofanti2021functional} also propose an alternative choice for step ($iii$), which relies on a studentized version of the functional residual in \eqref{eq_fof_res}. This approach is particularly suitable when the reference dataset is relatively small or when covariate shifts are present.
The studentized functional residual is defined as 
\begin{equation*}
e_{\text{stu}} (t) = \frac{Y (t) - \hat{Y}(t)}{\left(v_{\varepsilon} (t) + \bm \psi^Y(t)^T \bm \Sigma_{\varepsilon} \bm \psi^Y(t) \sum_{l=1}^L (\xi^X_{l})^2 / \eta^X_{l} \right)^{1/2}}, \quad t \in \mathcal T, 
\end{equation*}
where  $\bm \psi^Y = (\psi_{1}^Y, \dots, \psi_{M}^Y)^T$, $ v_{\varepsilon}^2$ is the residual variance  and $\bm \Sigma_{\varepsilon} $ is the  covariance matrix of $\epsilon_{m}$ in \eqref{eq_fof_simplified}.
Then, the FRCC monitors the residual using the strategy based on Hotelling’s $T^2$ and $SPE$ statistics, as described in Section \ref{sec_monitoring}.

A particular instance of the FRCC was considered by \cite{capezza2020control}, which focuses on a scalar quality characteristic. In this setting, the residual is also a scalar variable, and control limits are determined under the assumption of normality. Moreover, the selection of the retained principal components in the regression model is guided by prediction performance using the prediction sum of squares statistic.
\cite{wittenberg2025covariate} consider another instance of the FRCC framework by incorporating the general class of functional additive mixed models \citep{scheipl2015functional} in step ($i$). Their work primarily focuses on the case of a univariate functional covariate and adopts a concurrent regression model, in which the response at a given time depends on the value of the covariate at the same time point. 
Another application of the FRCC method is presented in \cite{capezza2022functional}.

\section{Robust Profile Monitoring in the Presence of Outliers}
\label{sec_out}
The monitoring framework described in Section \ref{sec_generalstrategy} requires a reference dataset that is assumed to represent the IC state of the process. This dataset is used in Phase I for both parameter estimation and the calculation of control limits.
However, in practice, such dataset is often contaminated with outliers, that is, observations deviating from the typical pattern of the data. Anomalies may arise from various sources, including measurement errors, sensor malfunctions, or rare and unexpected events. It is well known that control charts are highly sensitive to outliers in the reference dataset, as their presence can inflate control limits and significantly reduce the monitoring performance  \citep{vargas2003robust,chenouri2009multivariate}.

\begin{figure}
\begin{center}
\includegraphics[width=0.45\textwidth]{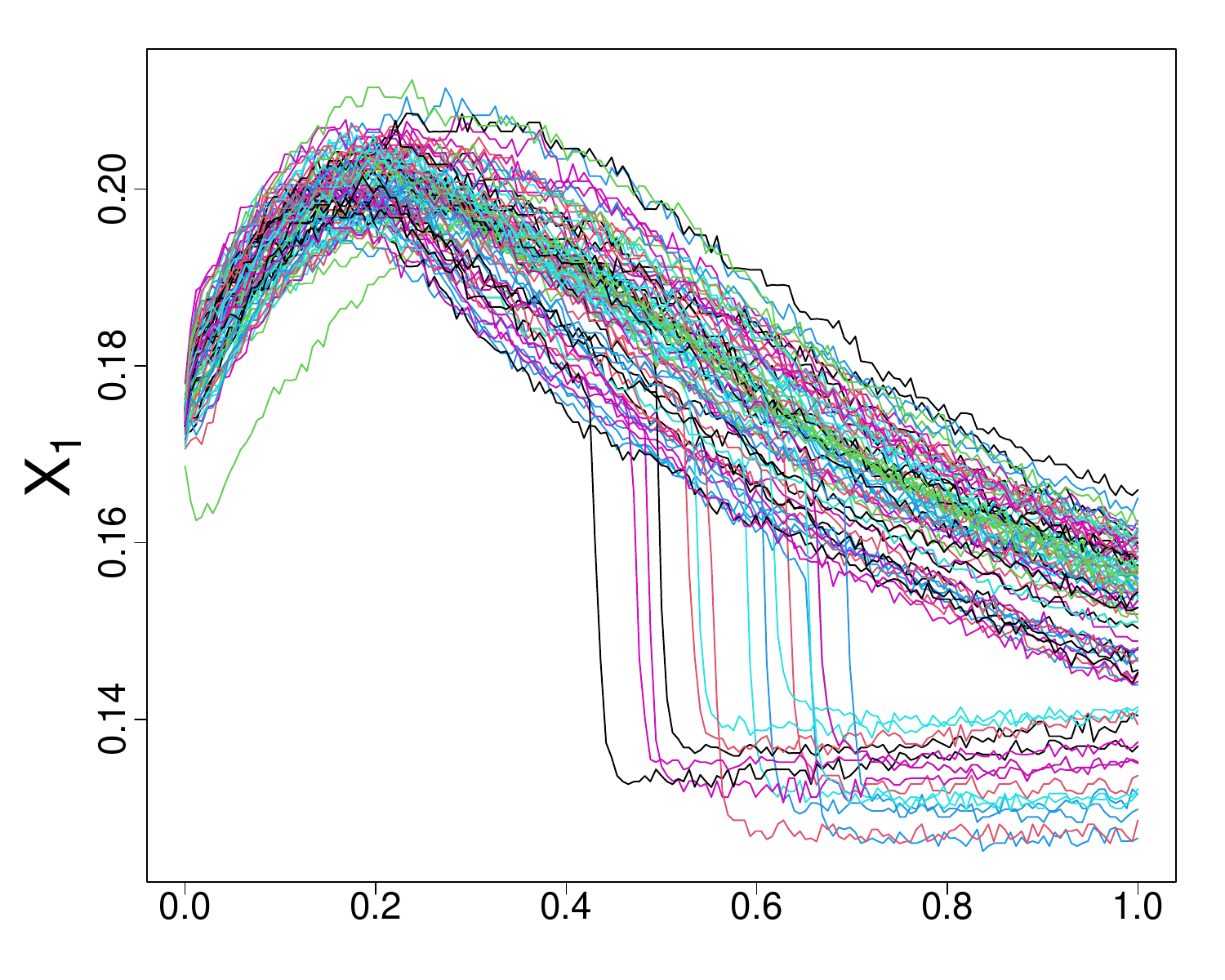}\includegraphics[width=0.45\textwidth]{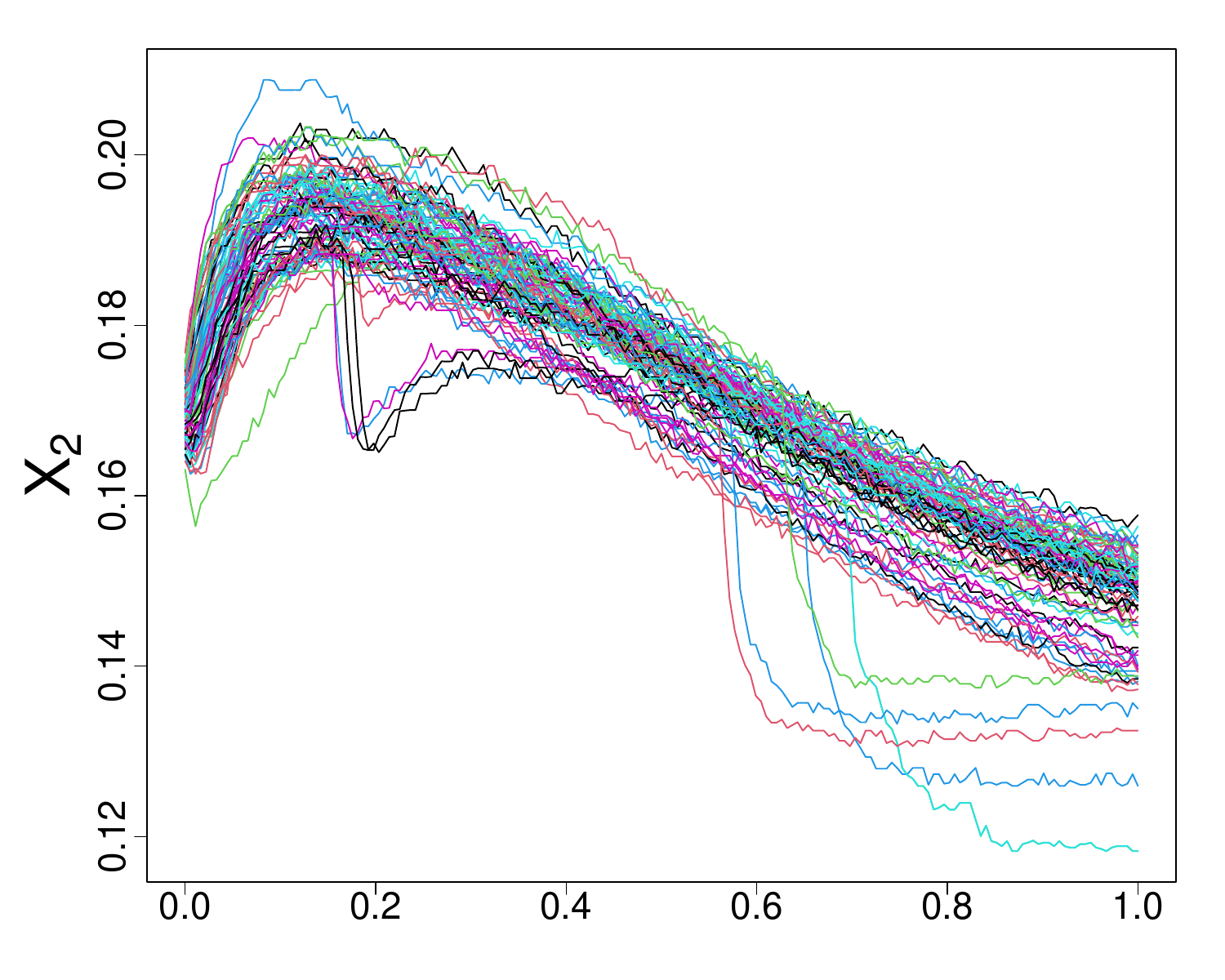}\\\includegraphics[width=0.45\textwidth]{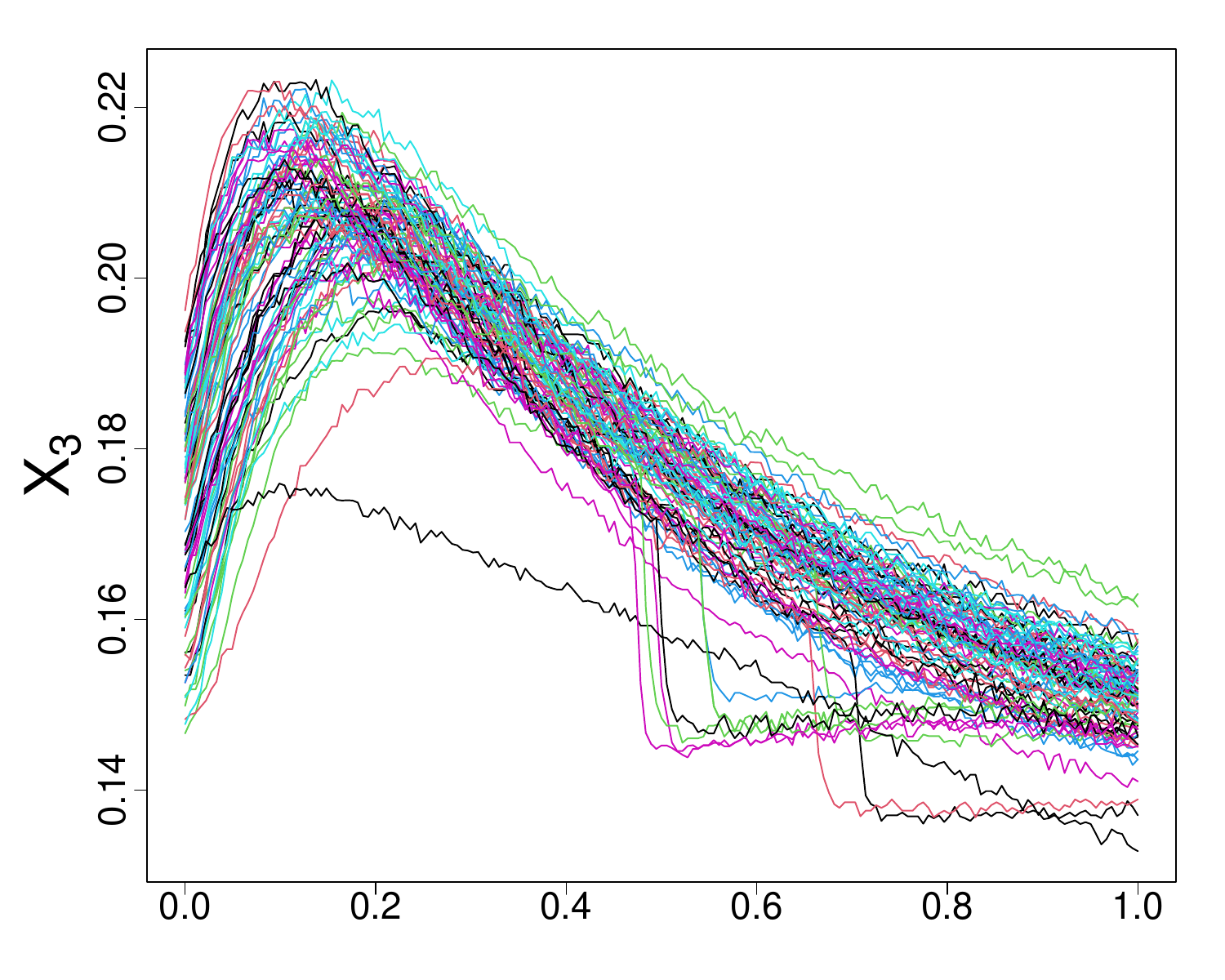}\includegraphics[width=0.45\textwidth]{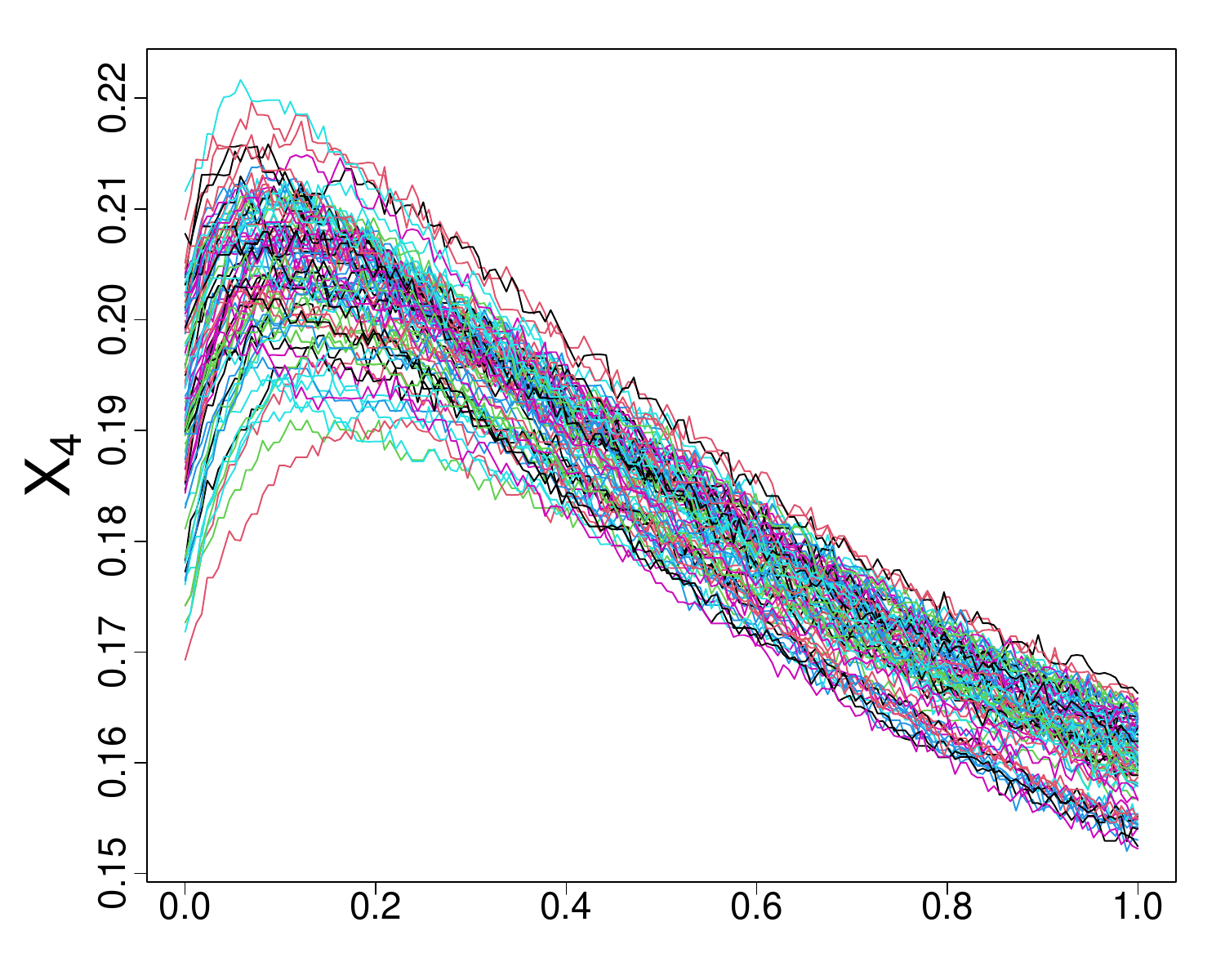}
\caption{A sample of 100 dynamic resistance curves acquired during a resistance spot welding process in an automotive application. Each panel corresponds to a different welding point.}
\label{fig_drc}
\end{center}
\vspace{-1cm}
\end{figure}

Although highly relevant in practice, this problem has received limited attention in the profile monitoring literature. A solution is proposed by \cite{capezza2024robust}.
This approach is called the robust multivariate functional control chart (RoMFCC) and is a Phase II profile monitoring method for multivariate functional data that is robust, i.e., outlier resistant, to the presence of both casewise and componentwise outliers in the reference sample.
A casewise outlier refers to an entire observation that deviates from the majority of the data. In contrast, a componentwise outlier refers to contamination that affects only certain components  of an observation, independently of the others. This form of contamination is particularly challenging in high-dimensional settings, as it can propagate across components and severely compromise traditional robust methods \citep{alqallaf2009propagation}.
As an illustration, Figure \ref{fig_drc} displays a severly contaminated sample of dynamic resistance curves  acquired during a resistance spot welding  process in an automotive application, where each panel corresponds to a specific welding point on an item produced by the same machine.

The RoMFCC is structured into four main elements:
($i$) a functional filter is used to identify functional componentwise outliers, which are then replaced by missing values to prevent contamination of the subsequent analysis;
($ii$) a robust multivariate functional data imputation method is applied to reconstruct the missing components in a way that preserves the overall structure of the data;
($iii$) a casewise robust dimensionality reduction technique, based on robust multivariate functional principal component analysis (RoMFPCA), is used to perform dimensionality reduction while down-weighting casewise outliers;
and ($iv$) a monitoring strategy, based on Hotelling’s $T^2$ and $SPE$ statistics, is employed.

The RoMFPCA is a  modification of the MFPCA method described in Section \ref{sec_MFPCA}.
Specifically, the eigenfunctions and eigenvalues of the covariance of the multivariate functional quality characteristic $\bm{X} = \left(X_{1},\dots,X_{p}\right)^T$ are estimated through robust principal component analysis, implemented by applying the ROBPCA method of \cite{hubert2005robpca} to $\bm{W}^{1/2}\widetilde{\bm{c}}$. This approach is computationally efficient and explicitly designed to achieve high breakdown performance in high-dimensional settings, thereby ensuring reliable estimation even in the presence of substantial casewise contamination.

The functional filter in ($i$) considers the  distances
\begin{equation*}
D_{k}^{\text{fil}}=\sum_{l=1}^{L^{\text{fil}}} 
    \frac{(\xi_{kl}^{\text{fil}})^2}{\eta_{kl}^{\text{fil}}}, \quad k=1,\dots,p,
\end{equation*}
where the  scores $\xi_{kl}^{\text{fil}}= \int_{\mathcal{T}}Z_{k}(t) {\psi}_{kl}^{\text{fil}}(t)dt$, eigenvalues $\eta_{kl}^{\text{fil}}$, principal components ${\psi}_{kl}^{\text{fil}}$, and standardized observations $Z_{k}$ of ${X}_{k}$ are obtained by applying the RoMFPCA  for each $k$.
In this case, RoMFPCA is used for representing distances among $X_{k}$'s and not for performing dimensionality reduction, thus $L^{\text{fil}}$ is set sufficiently large to capture a large fraction of total variability.
For each $k$, the proportion of flagged outliers is defined by 
\begin{equation*}
d_n = \sup_{x \geq \eta} \left\{ G(x) - G_n(x) \right\}^{+},
\end{equation*}
where $\left\{\cdot  \right\}^{+}$ denotes the positive part, $G_n$ is the empirical cumulative distribution function of $D_k^{\text{fil}}$, and $G$ is the reference CDF of $D_k^{\text{fil}}$, which is assumed to follow a chi-squared distribution with $L^{\text{fil}}$ degrees of freedom. The threshold $\eta$ is defined as $\eta = G^{-1}(\alpha)$, with $\alpha = 0.95$.
The proportion $d_n$ of observations with the largest functional distances $D_{k}^{\text{fil}}$ are identified as functional componentwise outliers, and the corresponding outlying components are replaced with missing values. 
On top of the univariate filter, a bivariate filter can also be applied to account for the correlation among components. 
This approach aims to identify outlying bivariate functional observations and is particularly useful for detecting moderately contaminated components that the univariate filter may not have flagged.

Then, step ($i$) outputs $n$ realizations $\bm{X}_i = \left(X_{i1}, \dots, X_{ip}\right)^T$, for $i = 1, \dots, n$, of the multivariate functional quality characteristic $\bm{X}$, with some missing components.  
The data imputation step ($ii$) starts from the realization $\bm{X}_{\underline{i}} = ((\bm{X}_{\underline{i}}^{\text{m}})^T, (\bm{X}_{\underline{i}}^{\text{o}})^T)^T$ having the smallest number $s$ of missing components, which is arranged such that the first $s$ missing components are in the vector $\bm{X}_{\underline{i}}^{\text{m}}$, while the remaining $p-s$ observed ones are in $\bm{X}_{\underline{i}}^{\text{o}}$.
Let us consider its standardized version $\bm{Z}_{\underline{i}}$, which is uniquely identified via \eqref{eq_fk} by a $(Kp)$-dimensional coefficient vector $\bm{c}_{\underline{i}} = \left((\bm{c}_{\underline{i}}^{\text{m}})^T, (\bm{c}_{\underline{i}}^{\text{o}})^T\right)^T$, corresponding to the basis expansions of its missing and observed components, respectively, where $K$ denotes the number of basis functions, identical for each component.
The imputation of $\bm{X}_{\underline{i}}^{\text{m}}$ reduces to the imputation of the corresponding coefficient vector $\bm{c}_{\underline{i}}^{\text{m}}$, which is obtained by minimizing \begin{equation*}
\sum_{l=1}^{L^{\text{imp}}} \frac{(\xi_{\underline{i}l}^{\text{imp}})^2}{\eta_{l}^{\text{imp}}},
\end{equation*}
where $\xi_{\underline{i}l}^{\text{imp}}$ are the scores associated with $\bm{Z}_{\underline{i}}$ and the principal components $\bm{\psi}_{l}^{\text{imp}}$ and eigenvalues $\eta_{l}^{\text{imp}}$, obtained by applying RoMFPCA to the subset of realizations without missing components. The number of components $L^{\text{imp}}$ is selected to be sufficiently large to explain a pre-specified proportion of the total variability.
Then, $\bm{c}_{\underline{i}}^{\text{m}}$ is estimated as
\begin{equation*}
\hat{\bm{c}}_{\underline{i}}^{\text{m}}= -\left( \bm{C}^{\text{m,m}}\right)^{\dagger}\bm{C}^{\text{m,o}}\bm{c}_{\underline{i}}^{\text{o}},
\end{equation*}
where $\bm{C}^{\text{m,m}}$ and $\bm{C}^{\text{m,o}}$ correspond to the upper-left and upper-right blocks of size $Ks \times Ks$ and $K(p-s) \times K(p-s)$, respectively, of the matrix $\bm{W}\bm{B}\bm{\Lambda}^{-1}\bm{B}^T\bm{W}$, and $(\cdot)^{\dagger}$ denotes the Moore-Penrose inverse.
The matrix $\bm{W}$ is defined as in Section~\ref{sec_MFPCA}, the columns of the matrix $\bm{B}$ contain the basis coefficients of the principal components $\bm{\psi}_{l}^{\text{imp}}$, and $\bm{\Lambda}=\diag(\eta_{1}^{\text{imp}},\dots,\eta_{L^{\text{imp}}}^{\text{imp}})$. 
To overcome the bias issue typical of deterministic imputation approaches, $\hat{\bm{c}}_{\underline{i}}^{\text{m}}$ is further perturbed by random noise.  
The entire imputation procedure described for $\bm{X}_{\underline{i}}$ is repeated iteratively until all realizations with missing components have been imputed. 

Once the data have been cleaned and imputed, the final step ($iv$) of the RoMFCC applies the monitoring framework for the multivariate functional quality characteristic $\bm{X}$, based on Hotelling’s $T^2$ and $SPE$ statistics, which is described in Section \ref{sec_monitoring}. However, since the presence of outliers prevents a reliable nonparametric estimation, the control limits are instead calculated parametrically, relying on the assumption of normality of the principal component scores.

\section{Real-time Monitoring of Functional Data}
\label{sec_FRTM}
\begin{figure}[h]
\centering \includegraphics[width=0.45\textwidth]{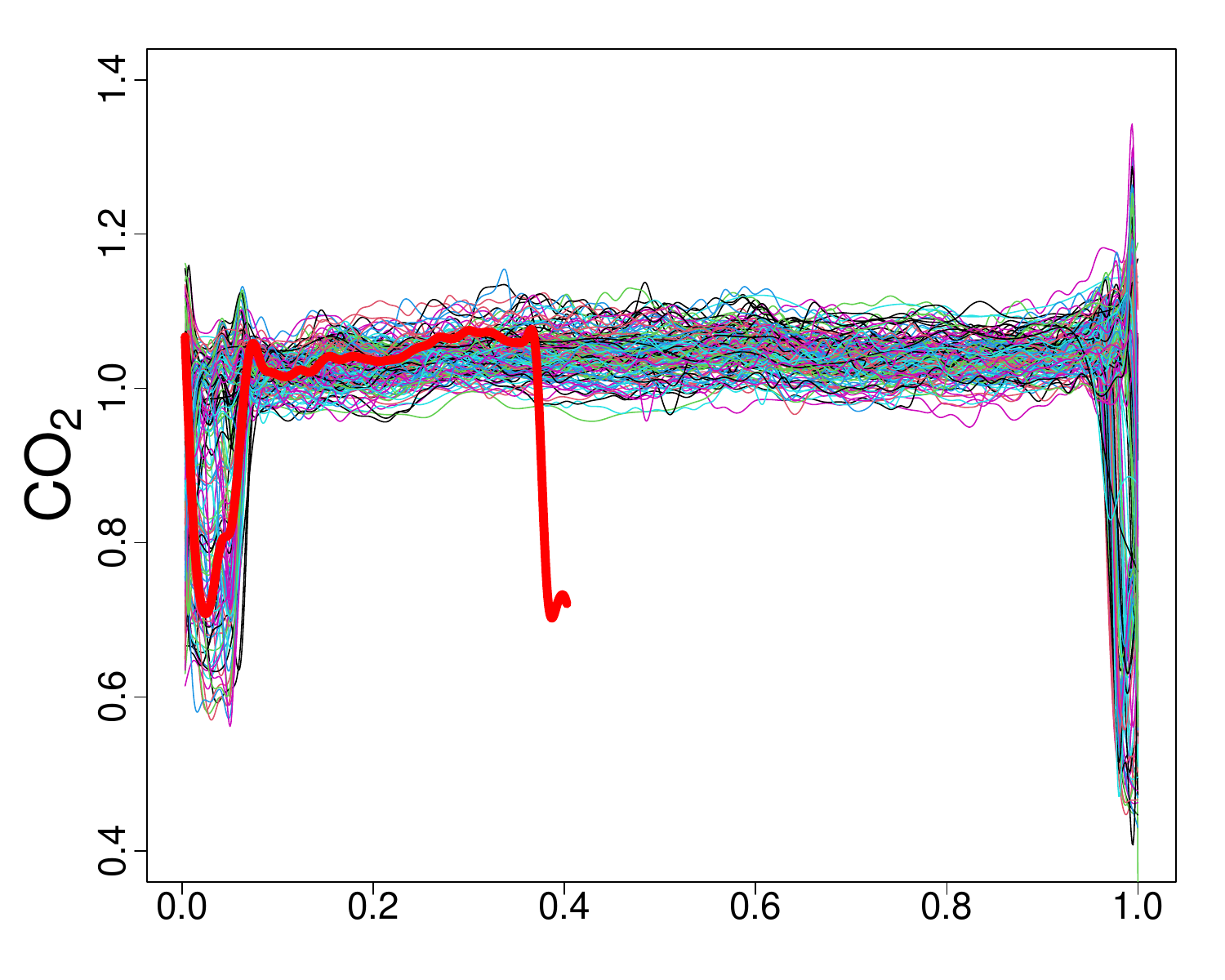}

\caption{\label{fig_real} CO\textsubscript{2} emissions profile observed up to a certain point during an ongoing voyage overlaid on the Phase I data from the ship operating condition monitoring example in Section \ref{sec_exam}.
}
\end{figure}

The profile monitoring methods presented so far consider the case of fully observed functional data. That is, they assess the presence of special causes once the functional quality characteristic has been completely observed.
In practice,  assessing the presence of special sources of variations in real time is also of great interest to significantly improve the monitoring effectiveness. In this setting, the attribute real-time stands for \textit{based on a functional quality characteristic partially observed up to the current domain point}.
This idea is illustrated in Figure \ref{fig_real}, which shows the CO\textsubscript{2} emissions profile observed up to a certain point during an ongoing voyage, overlaid on the Phase I data from the illustrative example in Section \ref{sec_exam}. In this case, it would be particularly valuable to detect the OC condition before the voyage is completed.
During the real-time monitoring phase, the functional quality characteristic, which is observed only up to a given point in its domain, must be compared to an appropriate reference distribution that reflects the IC  process up to that same point. However, identifying such a reference distribution is a non-trivial task, particularly because many processes exhibit phase variation, which refers to the lateral displacement of curve features (e.g., shifts in peaks or valleys along the domain), as opposed to amplitude variation, which concerns changes in the vertical scale of those features.

The functional real-time monitoring  (FRTM) method of \cite{centofanti2025real} allows monitoring a univariate functional quality characteristic in real time by taking into account both the phase and amplitude components identified through a real-time alignment procedure. 
FRTM consists of three main elements: ($i$) the registration of the partially observed functional data to the appropriate reference curve;
($ii$) the dimensionality reduction through a  mixed functional principal
component analysis; ($iii$) the joint monitoring in the reduce space of the phase and amplitude components through the application of Hotelling’s $T^{2}$  and $SPE$ control charts described in Section \ref{sec_monitoring}.

The curve registration problem in ($i$) is formalized as follows.
Let $ n $ functional observations $ X_i $, $ i=1,\dots, n$, be a random sample of functions whose realizations belong to $ L^2\left(\mathcal{D}_{X_i}\right) $. 
  Let $ X_i $ follow the general model
\begin{equation*}
X_i\left(x\right)=g_i\left(h_i^{-1}\left(x\right)\right), \quad x \in \mathcal{D}_{X_i},  i=1,\dots,n, 
\end{equation*}
where $ h_i: \mathcal{D}_{h_i}\rightarrow \mathcal{D}_{X_i}$ are strictly  increasing square integrable random functions, named warping functions, which are  defined on the compact domain $ \mathcal{D}_{h_i}=\left[a_{h_i},b_{h_i}\right]\subset \mathbb{R} $ and capture the phase variation by mapping the absolute time $t$ to  the observation time $x$, whereas, $ g_i$ are functions in $ L^2\left(\mathcal{D}_{h_i}\right) $, named amplitude functions,  which capture the amplitude variation.
The open-end/open-begin functional dynamic time warping (OEB-FDTW) 
estimates the warping function $ h_i $ by solving the following variational problem
\begin{equation}\label{eq_fdtw}
	\arginf_{h_i: \mathcal{D}_{h_i}\subseteq \mathcal{D}_{Y}\rightarrow\mathcal{D}_{X_i},\alpha_i\in[0,1]}\frac{1}{|\mathcal{D}_{h_i}|}\int_{\mathcal{D}_{h_i}}	\left[\widetilde{F}(t)\right.\left.+\lambda_i R\left(h'_i(t)-T_i/T_0\right)\right]dt,
\end{equation}
with 
\begin{equation*}
	\widetilde{F}(t)=\alpha_i^2\left(\frac{Y\left(t\right)}{||Y||}-\frac{X_i\left(h_i\left(t\right)\right)}{||X_i||}\right)^2+(1-\alpha_i)^2\left(\frac{Y'\left(t\right)}{||Y'||}-\frac{h'_i(t)X'_i\left(h_i\left(t\right)\right)}{||X'_i||}\right)^2\quad  t\in\mathcal{D}_{h_i},
\end{equation*}
where  $ R(u) $ is $u^2 $ if $ s^{min}\leq u+T_i/T_0\leq s^{max} $ or $ \infty $ otherwise, $ s^{min} $ and $ s^{max} $ are the minimum and maximum allowed values of the first derivative of $ h_i $,  $ ||f|| $ is the sup-norm of $ f $, and $ |\mathcal{D}_{h_i}|$ is the size of $ \mathcal{D}_{h_i} $. The parameter  $ \alpha_i $ tunes the dependence of the alignment on the amplitude of the curves $ Y(t) $ and $ X_i(h_i(t)) $,  and their first derivatives. Here, $Y \in L^2(\mathcal{D}_Y)$ denotes the template curve used for alignment.
The second right-hand term in Equation \eqref{eq_fdtw}, with the smoothing parameter $ \lambda_i\geq0 $, aims to penalize too irregular warping functions by constraining their first derivative to be not too steep and to lie between $ s^{min} $ and $ s^{max} $, with $T_i=|\mathcal{D}_{X_i}|$ and $T_0=|\mathcal{D}_{Y}|$. 
The OEB-FDTW method does not impose boundary constraints and allows the warping function to take values different from the boundary values of $\mathcal{D}_{X_i}$ at both the beginning and end of the process. The variational problem in \eqref{eq_fdtw} is solved using dynamic programming for a fixed value of $\alpha_i$, while the optimal $\alpha_i$ is selected through grid search. The smoothing parameter $\lambda_i$ is chosen based on the average curve distance ($ACD$) criterion, and the template function $Y$ is selected through the Procrustes fitting procedure. This iteratively registers all sample curves to a template, which is initially chosen as the average or a representative curve, and updates it as the mean of the aligned curves until convergence.

The second step ($ii$) applies mixed functional principal component analysis (mFPCA) to reduce the dimensionality of the pairs $\left(X^*_i, h_i\right)$ for $i = 1, \dots, n$, where $X^*_i(t) = X_i(h_i(t))$ is the aligned version of $X_i$, defined over the domain $\mathcal{D}_{h_i}$. 
To circumvent the fact that the space of warping functions is neither closed under addition nor equipped with a natural scalar product, the centered log-ratio ($clr$) transformation is applied to the first derivative of each warping function $h_i$. Since the domain $\mathcal{D}_{X_i}$ of each warping function may differ across observations, these functions are first mapped to a common reference domain as
\begin{equation*}
\widetilde{h}_i\left(t\right)=\frac{	h_i\left(t\right)-F_{0i}}{F_{1i}-F_{0i}}, \quad t\in \mathcal{D}_{h_i},
\end{equation*} 
which is defined on the range $ \left[0,1\right] $,
where $ F_{0i} $ and $ F_{1i} $ are the values of $ h_i $ at the boundaries of $ \mathcal{D}_{h_i} $. Then, the  $clr$ transformation  is applied as
\begin{equation*}
	v_i(t)=clr(\widetilde{h}'_i)(t)=\log\left(\widetilde{h}'_i(t)\right)-\frac{1}{|\mathcal{D}_{h_i}|}\int_{\mathcal{D}_{h_i}}\log\left(\widetilde{h}'_i(t)\right)dt, \quad t\in \mathcal{D}_{h_i}.
\end{equation*}
Then, the MFPCA methods, described in Section \ref{sec_MFPCA}, is applied to the vector $ Z_i=(X^*_i,v_i, F_{0i}, \widetilde{F}_{1i})$, which is defined in $L^2\left(\mathcal{D}_{h_i}\right)\times L^2\left(\mathcal{D}_{h_i}\right)\times\mathbb{R} \times\mathbb{R}$, where $ \widetilde{F}_{1i}=log(F_{1i}-F_{0i}) $ is introduced to respect the constraint $ F_{1i}>F_{0i} $. 
The components  $v_i$, $F_{0i}$, and $\widetilde{F}_{1i}$ jointly capture information about the phase component and may exhibit different variability compared to $X_i^*$. Therefore, before representing $X_i^*$ and $v_i$ as a linear combination of basis functions, each component is appropriately standardized to ensure a proper balance in the contribution of amplitude and phase variation during the dimensionality reduction step. Moreover, to address the fact that each pair $(X^*_i, h_i)$ may be defined on different domains due to the open-begin/open-end nature of the registration method, the problem is reformulated as a missing data imputation task, where the missing parts are imputed using a suitably shifted template curve to ensure continuity.

In Phase I, a preprocessing step is performed in which the common monitoring domain $\mathcal{D}_{m}$ is selected. The Procrustes fitting procedure and the $ACD$ criterion are then used to determine an appropriate template function $Y$ and the smoothing parameter $\lambda$.
Next, in the real-time registration step, the IC functional observations are aligned to the template function, as if they were observed up to each specific point in $\mathcal{D}_{m}$. Once all IC observations are registered in real time, mFPCA is applied for each $t \in \mathcal{D}_Y$, based on the sample $\left(X^*_{i,t}, h_{i,t}\right)$.
Then, the Hotelling's $T^2$ and $SPE$ statistics are computed as in Section \ref{sec_monitoring}, along with their corresponding control limits $CL^{T^2}_x$ and $CL^{SPE}_x$, for each observation and  $x \in \mathcal{D}_m$.
In Phase II, for a new incoming observation $X_{\text{new},x}$ of the functional quality characteristic observed up to $x \in \mathcal{D}_m$, the values $T^2$ and $SPE$ are computed. An OC signal is issued if either $T^2$ or $SPE$ exceeds its respective control limit, $CL^{T^2}_x$ or $CL^{SPE}_x$.

Another possible real-time monitoring strategy for functional data is proposed by \cite{capezza2020control} in the context of monitoring CO\textsubscript{2} emissions in a maritime application. However, this approach does not account for potential phase variation and is specifically tailored to the particular application under consideration, thereby lacking of generality. Another approach for real-time monitoring is proposed by \cite{austin2024detection}.

\section{Adaptive Methods for Profile Monitoring}
\label{sec_adaptive}
The reference monitoring framework in Section \ref{sec_generalstrategy}  can be viewed as a functional extension of the Shewhart  monitoring strategy, which is the most widely used paradigm in statistical process monitoring. This approach assesses the process state using only the information available at the current time point, without incorporating the temporal evolution of previous observations. While this strategy is well-suited for detecting large shifts, it tends to be less effective for identifying small but persistent changes. 
Another limitation of this strategy concerns the procedure used to select the parameters required for its implementation. The data smoothing procedure described in Section \ref{sec_smoothing} depends on the selection of smoothing parameters $\lambda_k$, and the MFPCA approach in Section \ref{sec_MFPCA} requires determining the number of retained principal components $L$. These parameters are typically selected according to criteria that focus on minimizing estimation or prediction errors, such as cross-validation or generalized GCV for smoothing, and by retaining a number of principal components sufficient to explain a specified proportion of the total variability in the case of dimensionality reduction. However, it is well recognized in the literature that the parameter values optimal for model estimation are not necessarily optimal for monitoring purposes.
These two issues share a common origin, namely that the monitoring strategy is not explicitly tailored to the characteristics of the OC distribution.
As a consequence, monitoring performance may degrade in real-world applications,  particularly when the actual OC behavior deviates significantly from the implicit assumptions made.
In this section, two methods are presented that address the aforementioned limitations and fall under the broader category of adaptive procedures. From a broad perspective, these approaches aim to improve monitoring performance by explicitly adapting to the unknown OC distribution, enhancing sensitivity to a wider range of deviations from the IC process.

\cite{capezza2025adaptive} propose an adaptive multivariate functional exponentially weighted moving average (AMFEWMA) monitoring scheme for the SPM of multivariate functional quality characteristics. This approach addresses the limitations of the Shewhart monitoring strategy by assessing the current state of the process through a statistic computed as a weighted average of the current observation and the corresponding statistic from the previous time point.
The weighted average is determined by a single weighting parameter, which is adaptively selected to better perform under a wide range of OC conditions.

Let us consider a sequence of realizations of the quality characteristic \sloppy $\bm X_i = (X_{i1}, \dots, X_{ip})^T$, $i=1,2,\dots$, which are assumed, without loss of generality, to be mean centred.
The AMFEWMA detects changes in the process by using the statistic
\begin{equation}
     \label{AMFEWMA_statistic}
         {\bm Y}_{i}(t) = 
         ({\bm I} - {\bm\Lambda}_{i}(t)){\bm Y}_{i-1}(t) + {\bm \Lambda}_{i}(t)  {\bm X}_{i}(t),
         \quad t \in \mathcal{T}, \quad i=1,2,\dots,
     \end{equation}
where ${\bm \Lambda}_{i}(t) = \diag\left(w(E_{i1}(t)), \dots, w(E_{ip}(t))\right)$, with
$w(E_{ik}(t)) = \eta(E_{ik}(t)) / E_{ik}(t)$, and $\eta(E_{ik}(t))$ denotes a score function evaluated at $E_{ik}(t)$. Moreover,
$ {\bm E}_i (t) = ( E_{i1} (t),\dots,  E_{ip} (t))^T = {\bm X}_{i}(t) - {\bm Y}_{i-1} (t)$.
The AMFEWMA statistic ${\bm Y}_{i} $ results in a weighted average of the current observation ${\bm X}_{i}$ and the charting statistic value at the previous time point ${\bm Y}_{i-1}$, as in a conventional EWMA chart, but with weights changing over time. 
The score function $\eta$  must be strictly increasing, odd ($\eta(t) = -\eta(-t)$), and such that $\eta(E_{ik}(t))$ is close to $\lambda E_{ik}(t)$ (resp. to $E_{ik}(t)$) when $|E_{ik}(t)|$ is small (resp. large).
An appropriate choice is
\begin{equation*}
\widetilde{\eta}(E_{ik}(t)) = 
    \begin{cases}
    E_{ik}(t)+(1-\lambda) C_k(t) & \text{if } E_{ik}(t) < -C_k(t),\\
    \lambda E_{ik}(t) & \text{if } | E_{ik}(t)|\leq C_k(t),\\
    E_{ik}(t)-(1-\lambda) C_k(t) & \text{if } E_{ik}(t)>C_k(t),
\end{cases} \quad t \in \mathcal T,
\end{equation*}
where $\lambda \in (0,1]$ and $C_k(t) = k \sigma_{j}(t)$, with $k>0$, and $\sigma_1 (t), \dots, \sigma_p (t)$ are the standard deviation function of  $\bm X_i$.
When $|E_{ik}(t)|$ tends to zero, the AMFEWMA performs like a classical EWMA with parameter $\lambda$, otherwise it behaves as a Shewhart control chart, therefore adapting to the unknown OC distribution.
Then, the monitoring statistic is defined as the Hotelling's $T^2$ statistic in \eqref{eq_T2}, computed by applying the MFPCA procedure described in Section \ref{sec_MFPCA} to  ${\bm Y}_{i}$.
In Phase I and for fixed values of $\lambda$ and $k$, a bootstrap procedure is applied to the reference sample to compute the control limit such that a pre-specified average run length (ARL) is achieved, i.e., the expected number of observations required to signal an OC condition. Additionally, the optimal values of the parameters $\lambda$ and $k$ are selected through a dedicated algorithm.

To address the issue that optimal parameters for model estimation are generally not optimal for monitoring, \cite{centofanti2025adaptive} present a new Phase II monitoring framework, referred to as adaptive multivariate functional control chart (AMFCC),  that chooses the parameters by adapting to the unknown distribution of the OC observations.
The AMFCC method is largely built upon the reference monitoring framework outlined in Section \ref{sec_generalstrategy}. However, unlike this approach, it does not fix the smoothing parameters and the number of retained components based on estimation criteria. Instead, it considers multiple combinations of these parameters and integrates the corresponding Hotelling $T^2$-type statistics into a single, aggregated monitoring statistic.
To avoid a prohibitive number of parameter combination the smoothing parameters in \eqref{eq_smootspline2} are set as
\begin{equation*}
	\lambda_k=\lambda\frac{w_k}{\sum_{i=1}^pw_i}, \quad k=1,\dots,p,
\end{equation*}
where $w_k=1/((\bm{c}^0_{k})^T\bm{R}_{k}^{(m)}\bm{c}^0_{k})^{-1}$, with $\bm{c}^0_{k}$ the solution of the optimization problem in Equation \eqref{eq_smootspline2} with $\lambda_1=\dots=\lambda_p=\lambda$. Thus, the unique parameter of the data smoothing is $\lambda$. 
The AMFCC considers the following Hotelling $T^2$-type statistics
\begin{equation*}
	T^2_{\lambda, L}=
    \sum_{l=1}^{L}\frac{\xi_{l;\lambda}^2}{\eta_{l;\lambda}},
\end{equation*}
where $\xi_{l;\lambda}$ and $\eta_{l;\lambda}$ denote the scores and eigenvalues, respectively, obtained by applying the data smoothing and MFPCA procedures described in Sections \ref{sec_smoothing} and \ref{sec_MFPCA}, using fixed values of  $\lambda$.
Let us indicate $T^2_{\lambda, L}$ with $T^2_{t} $.
The main idea behind AMFCC is that the null hypothesis $H_0$, stating that the process is IC, can be decomposed into a set of sub-hypotheses $H_{0t}$ for $t = 1, \dots, T$, such that $H_0$ holds if and only if all $H_{0t}$ are simultaneously true. Each sub-hypothesis $H_{0t}$ is tested using the statistic $T^2_t$, where $T$ denotes the number of parameter combinations considered.

The statistic $T^2_t$ are combined  in an overall monitoring statistic $T^2$ by considering their associated $p$-values $p_{1},\dots,p_{T}$ and the combining function $\Theta$ as
\begin{equation}\label{eq_theta}
	T^2=\Theta(p_{1},\dots,p_{T}),
\end{equation} 
where $\Theta:(0,1)^{T}\rightarrow \mathbb{R}$ is a continuous and non-increasing. 
The use of the $p$-values ensures that the combination takes place on a common scale. 
Then, the AMFCC  triggers a signal if $T^2$ is above a control limit, which is obtained as described in Section \ref{sec_monitoring}.
Common choices for $\Theta$ are  the Fisher omnibus \citep{fisher1992statistical} and Tippett \citep{tippett1931methods} combining functions, which are defined respectively as 
\begin{equation*}
T^2_{F}=-2\frac{\sum_{t=1}^{T}\log(p_{t})}{T}, \quad T^2_{T}=-2\log \left(\min_{1\le t \le T}p_{t}\right).
\end{equation*}
The AMFCC also provides a diagnostic tool that, building on the same idea of combining statistics, aggregates the contributions of each component to the $T^2$ and $SPE$ statistics in \eqref{eq_cont}, computed across different parameter combinations.
It is worth noting that the AMFCC framework is general and can be readily integrated into alternative monitoring schemes that involve parameter selection for their implementation.

\section{Additional Methods for Monitoring Functional Data}
\label{sec_additional}
While the approaches presented in the previous sections form the core of FDA-based profile monitoring, they do not exhaust the range of available methodologies. A review of additional existing FDA-based profile monitoring methods is now presented, thereby offering a more comprehensive description of the current state of the literature.

One of the building blocks of the reference monitoring framework presented in Section \ref{sec_generalstrategy} is the use of functional principal component analysis (FPCA) for dimensionality reduction.
FPCA has also been employed by \cite{yu2012outlier}, who propose a monitoring approach for univariate functional data that, similarly to the framework in Section \ref{sec_generalstrategy}, relies on FPCA to construct a Hotelling's $T^2$ statistic.
Control limits are derived based on asymptotic considerations under the assumption of Gaussianity.
FPCA is also used by \cite{grasso2016using}, who introduce a functional data monitoring framework that explicitly incorporates curve registration into the construction of the control chart. Similarly to the FRTM method in Section \ref{sec_FRTM}, this approach does not treat registration as a preprocessing step but jointly monitors amplitude and phase variation by applying FPCA to registered profiles and incorporating warping parameters into a monitoring scheme based on the Hotelling's $T^2$ and $SPE$ statistics. However, differently from FRTM, it is designed only for completely observed functional data. The problem of monitoring unaligned profiles is also addressed by \cite{zang2016unaligned}.
Other works that employ FPCA for dimensionality reduction include \cite{menafoglio2018profile} and \cite{scimone2022statistical}, where FPCA is adapted into the simplicial functional principal component analysis (SFPCA) to handle functional data represented as probability density functions. In both cases, the monitoring strategy relies on the simultaneous use of Hotelling's $T^2$ and $SPE$ statistics, with control limits empirically obtained from the reference data set.

\cite{paynabar2016change} propose a change-point detection approach for Phase I analysis of multivariate functional data, based on multichannel functional principal component analysis (mcFPCA). Differently from the MFPCA approach described in Section \ref{sec_MFPCA}, mcFPCA represents the multivariate functional data $\bm{X}=(X_1,\dots,X_p)^T$ as
\begin{equation*}
\bm{X}(t) = \sum_{l=1}^{\infty} \bm{\xi}_l \nu_l(t), \quad t \in \mathcal{T},
\end{equation*}
where the scores $\bm{\xi}_l$ are independently and identically distributed $p$-variate random variables with zero mean, and the princiapl components $\nu_l$ form an orthonormal basis in $L^2(\mathcal{T})$. Unlike MFPCA, mcFPCA assumes that all $p$ functional components share the same set of eigenfunctions. The relations among components are thus captured by the correlation structure of the multivariate scores $\bm{\xi}_l$.
The monitoring strategy employs a Hotelling’s $T^2$ statistic, which is constructed from the standardized difference in means before and after a potential change point projected onto the principal components obtained via mcFPCA.
 Using the same framework, \cite{wang2018thresholded} propose a modification of the Hotelling’s $T^2$ monitoring statistic to better handle scenarios in which the change occurs along unknown principal component directions. This is a realistic situation in practice, especially when a large number of principal components are retained. Specifically, components of the $T^2$ statistic below a certain threshold are set to zero, effectively filtering out directions that are unlikely to be affected by the change and that do not contain meaningful information for detection.  

\cite{ren2019phase} extend the approach of \cite{paynabar2016change} to Phase II monitoring. Specifically, they integrate the mcFPCA-based dimensionality reduction into an exponentially weighted moving average (EWMA) control scheme. An EWMA sequence is defined based on the mcFPCA scores $\bm{\xi}_{il}$, corresponding to the realizations $\bm{X}_i$, $i=1,2,\dots,$ of $\bm{X}$, as follows
\begin{equation}
\label{eq_ewmaren}
    \bm{\eta}_{il} = (1 - \lambda)\bm{\eta}_{i-1,l} + \lambda \bm{\xi}_{il}, \quad l = 1, \dots, L,
\end{equation}
where $\lambda \leq 1$ is the smoothing parameter and $L$ is the number of retained principal components. A signal is triggered if the monitoring statistic
\begin{equation*}
    Q_i = \frac{2 - \lambda}{\lambda} \sum_{l = 1}^{L} \bm{\eta}_{il}^T \bm{\Sigma}_l^{-1} \bm{\eta}_{il},
\end{equation*}
exceeds a threshold $CL > 0$. Here, $\bm{\Sigma}_l$ denotes the covariance matrix of the $l$-th score vector.
\cite{alqahtani2020multilevel} apply a specific instance of the method proposed by \cite{ren2019phase} to detect local changes in 3D topographic surfaces. In particular, they consider univariate profiles, and implement a Shewhart-type monitoring strategy by setting $\lambda = 1$ in \eqref{eq_ewmaren}.
\cite{zhang2015phase} describe a monitoring framework for Phase I that applies a change-point detection strategy to the residuals obtained from regressing each component of the multivariate functional data against the others. The method of \cite{jalilibal2025novel} also derives monitoring statistics based on the scores obtained from the  mcFPCA. 

Building on \cite{paynabar2016change}, \cite{zhang2018multiple} propose a systematic framework for online monitoring of high-dimensional streaming data in manufacturing systems. The approach includes a preprocessing phase in which data are synchronized to account for phase variation, and long-term drifts are removed using a fixed-effects model. A clustering algorithm is then applied to group functional components with similar features. Then, for each cluster, mcFPCA is used to obtain Hotelling's $T^2$ and $SPE$ statistics based on the EWMA sequences of the scores and residuals, similarly to \cite{ren2019phase}. Finally, a data-fusion strategy is employed to combine the cluster-level statistics into a single overall monitoring statistic.
\cite{yao2023adaptive} address the problem of the sampling constraint, where only a subset of the multivariate functional components can be observed at each time point. In the offline phase, fully observed multivariate functional data are used to perform mcFPCA, yielding empirical estimates of eigenfunctions and scores. During online monitoring, scores for the observable components are directly computed, while those for the unobserved components are imputed by sampling from the empirical distribution of historical MFPC scores. A multivariate cumulative sum (MCUSUM) statistic is then constructed using the completed score vectors. To determine which components to observe in the next step, an adaptive sampling strategy is employed, selecting the components associated with the largest elements of the MCUSUM statistic.
This approach is further extended in \cite{yao2024adaptive} to monitor partially observed images.

Another line of research avoids data-adaptive dimensionality reduction via functional principal components. Instead, this approaches work directly on the basis coefficients used to reconstruct functional data from discrete observations. 
When placed within the monitoring framework  described in Section \ref{sec_generalstrategy}, these methods operate directly on the coefficient vector $\bm{c}_k$ used to represent each component of $X_k$ as a linear combination of known basis functions. 
\cite{zhou2006spc} used wavelet transforms based on the Haar basis to represent univariate functional data. The resulting Haar coefficients are then treated as process features and monitored using a Hotelling's $T^2$ statistic. Similarly, \cite{chicken2009statistical} apply wavelet transforms for dimensionality reduction and use a likelihood ratio test (LRT) to detect possible change-points.
\cite{chang2010statistical} use wavelet denoising followed by B-spline modelling to extract control points from univaraite profile, which are then monitored via a Hotelling's $T^2$ statistic.
\cite{jeong2006wavelet} propose a monitoring statistic based on the differences between the wavelet coefficients of the observed and baseline signals, applying hard-thresholding to retain only the coefficients that significantly differ from zero.
The wavelet coefficients are modelled using random-effect and mixed-effects models in \cite{paynabar2011characterization} and \cite{jeong2018statistical}, and subsequently monitored through a LRT for change-point detection in the former, and an EWMA strategy for Phase II monitoring in the latter.
A Bayesian monitoring approach based on wavelet decompositions is proposed by \cite{shamp2021computationally}.
In \cite{colosimo2022complex}, B-spline coefficients are used to represent the profiles and are monitored using a Hotelling's $T^2$ statistic.
Another method for monitoring multiple correlated nonlinear profiles based on the B-spline coefficients extracted from each profile is proposed by \cite{chou2014monitoring}. 

Other approaches \citep{zhang2018weakly,wang2021sparse} perform dimensionality reduction by directly applying low-rank approximation methods to the discretely observed data $\lbrace Y_{k}(t_{kj}), t_{kj} \in \mathcal{T}, j = 1, \dots, n_k \rbrace_{k=1,\dots,p}$. However, these methods require the observations to lie on a common grid and do not explicitly account for the smoothness of functional data. In contrast, \cite{yan2018real} propose a spatio-temporal smooth sparse decomposition that explicitly models the functional nature of the data. The resulting estimated scores are then used to construct a likelihood ratio test statistic, which is incorporated into either a Shewhart-type or EWMA monitoring scheme.

A different perspective is taken by \cite{wu2022monitoring}, whose aim is to monitor the heterogeneity in the between-component relationships of multivariate functional data. This problem is reformulated as monitoring changes in the underlying graphical networks that represent interdependencies among components. Each component is first represented through scores extracted by applying univariate FPCA. A heterogeneous graphical model is then developed using these scores, and a Shewhart-type monitoring strategy is applied to the model residuals.
 Other FDA-based monitoring approachesd include \cite{liu2023covariate} and \cite{eslami2023statistical}.
\section{Conclusions and Suggestions for Future Research}
\label{sec_conclusion}
Profile monitoring is emerging as a relevant and increasingly important area in applied statistics.
Functional data analysis provides effective solutions to the challenges of statistical process monitoring  for profile data, overcoming many of the limitations associated with more classical profile monitoring approaches.

This review presents a range of FDA-based profile monitoring methods, beginning with a reference framework for monitoring multivariate functional data, which serves as the foundation for several recently developed approaches.
This review further discusses methods that incorporate additional functional covariates in the monitoring, a robust strategy designed to accommodate outlying observations, a real-time technique suited for scenarios in which the functional quality characteristic is only partially observed, and adaptive strategies that target the characteristics of the OC distribution.
A survey of additional FDA-based profile monitoring approaches is also included, providing a comprehensive overview of the current literature.

Although this review highlights the wide range of FDA-based approaches available to tackle many practical challenges, several important research opportunities and open problems remain. A central challenge arises from the fact that modern industrial systems generate increasingly high-frequency, high-dimensional data streams, often collected from multiple heterogeneous sensors. Scaling FDA-based profile monitoring methods to handle this complexity while ensuring both computational and statistical efficiency remains unresolved.

Some proposals have addressed phase variation, yet methods that can automatically and efficiently separate phase and amplitude components in multivariate profiles are still lacking. Misalignment remains a major challenge, as it can affect detection performance.

Another key limitation is that processes evolve dynamically, while most FDA-based monitoring methods rely on fixed parameters estimated from a fixed Phase I sample. Developing approaches that can learn adaptively from streaming data and update parameters in real time remains an open problem, particularly in start-up or small-batch settings where large Phase I data are unavailable.

The FRTM method (Section \ref{sec_FRTM}) assesses the presence of special causes of variation before the functional quality characteristic has been fully observed. This reflects a broader and important challenge involving situations where profiles are only partially observed, either due to missing measurements or sampling constraints imposed by limited resources.
Effective methods for handling these scenarios are still limited.

Robustness to outliers represents another critical issue that has received limited attention, even though anomalous observations can severely affect the analysis. Currently, the ROMFCC method (Section \ref{sec_out}) is the only approach that explicitly addresses this issue.

Beyond detection, diagnostic tools also require further development. Contribution plots (Section \ref{sec_monitoring}) provide insight into which variables trigger an alarm, but they represent only an initial step. Complementary techniques are needed to support interpretation and root-cause analysis. For instance, combining control charts with multivariate change-point estimators could help determine not only when a shift begins but also how long it persists.

Finally, while the \texttt{funcharts} package provides a valuable foundation,  further software development is needed to support the integration of FDA-based profile monitoring methods into practice. Highly optimized, real-time implementations that integrate seamlessly with industrial protocols and business systems are needed. Moreover, the scarcity of publicly available benchmark datasets further limits research progress.


\section*{Data Availability Statement}
Data sharing is not applicable to this article as no new data were created or analyzed in this study.

\section*{Disclosure Statement}
The authors report there are no competing interests to declare.

\bibliographystyle{apalike}
\setlength{\bibsep}{5pt plus 0.2ex}
{\small
\spacingset{1}
\bibliography{References}

\begin{thebibliography}{}

\bibitem[Alqahtani et~al., 2020]{alqahtani2020multilevel}
Alqahtani, M.~A., Jeong, M.~K., and Elsayed, E.~A. (2020).
\newblock Multilevel spatial randomness approach for monitoring changes in 3{D} topographic surfaces.
\newblock {\em International Journal of Production Research}, 58(18):5545--5558.

\bibitem[Alqallaf et~al., 2009]{alqallaf2009propagation}
Alqallaf, F., Van~Aelst, S., Yohai, V.~J., and Zamar, R.~H. (2009).
\newblock Propagation of outliers in multivariate data.
\newblock {\em The Annals of Statistics}, 37(1):311 -- 331.

\bibitem[Austin et~al., 2024]{austin2024detection}
Austin, E., Eckley, I.~A., and Bardwell, L. (2024).
\newblock Detection of emergent anomalous structure in functional data.
\newblock {\em Technometrics}, 66(4):614--624.

\bibitem[Capezza et~al., 2025a]{capezza2025adaptive}
Capezza, C., Capizzi, G., Centofanti, F., Lepore, A., and Palumbo, B. (2025a).
\newblock An adaptive multivariate functional {EWMA} control chart.
\newblock {\em Journal of Quality Technology}, 57(1):1--15.

\bibitem[Capezza et~al., 2022]{capezza2022functional}
Capezza, C., Centofanti, F., Lepore, A., Menafoglio, A., Palumbo, B., and Vantini, S. (2022).
\newblock Functional regression control chart for monitoring ship {CO}$_2$ emissions.
\newblock {\em Quality and Reliability Engineering International}, 38(3):1519--1537.

\bibitem[Capezza et~al., 2023]{capezza2023funcharts}
Capezza, C., Centofanti, F., Lepore, A., Menafoglio, A., Palumbo, B., and Vantini, S. (2023).
\newblock Funcharts: Control charts for multivariate functional data in {R}.
\newblock {\em Journal of Quality Technology}, 55(5):566--583.

\bibitem[Capezza et~al., 2025b]{capezza2025funcharts}
Capezza, C., Centofanti, F., Lepore, A., Menafoglio, A., Palumbo, B., and Vantini, S. (2025b).
\newblock funcharts: Functional control charts.
\newblock {\em R package version 1.7.0}.

\bibitem[Capezza et~al., 2021]{capezza2021functional}
Capezza, C., Centofanti, F., Lepore, A., and Palumbo, B. (2021).
\newblock A functional data analysis approach for the monitoring of ship {CO}$_2$ emissions.
\newblock {\em Gest{\~a}o \& Produ{\c{c}}{\~a}o}, 28:e152.

\bibitem[Capezza et~al., 2024]{capezza2024robust}
Capezza, C., Centofanti, F., Lepore, A., and Palumbo, B. (2024).
\newblock Robust multivariate functional control chart.
\newblock {\em Technometrics}, 66(4):531--547.

\bibitem[Capezza et~al., 2020]{capezza2020control}
Capezza, C., Lepore, A., Menafoglio, A., Palumbo, B., and Vantini, S. (2020).
\newblock Control charts for monitoring ship operating conditions and {CO}\textsubscript{2} emissions based on scalar-on-function regression.
\newblock {\em Applied Stochastic Models in Business and Industry}, 36(3):477--500.

\bibitem[Cardot et~al., 2003]{cardot2003spline}
Cardot, H., Ferraty, F., and Sarda, P. (2003).
\newblock Spline estimators for the functional linear model.
\newblock {\em Statistica Sinica}, 13:571--591.

\bibitem[Centofanti et~al., 2025a]{centofanti2025real}
Centofanti, F., Kulahci, M., Lepore, A., and Spooner, M.~P. (2025a).
\newblock Real-time monitoring of functional data.
\newblock {\em Journal of Quality Technology}, 57(2):135--152.

\bibitem[Centofanti et~al., 2021]{centofanti2021functional}
Centofanti, F., Lepore, A., Menafoglio, A., Palumbo, B., and Vantini, S. (2021).
\newblock Functional regression control chart.
\newblock {\em Technometrics}, 63(3):281--294.

\bibitem[Centofanti et~al., 2025b]{centofanti2025adaptive}
Centofanti, F., Lepore, A., and Palumbo, B. (2025b).
\newblock An adaptive multivariate functional control chart.
\newblock {\em Technometrics}, (just-accepted):1--31.

\bibitem[Chang and Yadama, 2010]{chang2010statistical}
Chang, S.~I. and Yadama, S. (2010).
\newblock Statistical process control for monitoring non-linear profiles using wavelet filtering and b-spline approximation.
\newblock {\em International Journal of Production Research}, 48(4):1049--1068.

\bibitem[Chenouri et~al., 2009]{chenouri2009multivariate}
Chenouri, S., Steiner, S.~H., and Variyath, A.~M. (2009).
\newblock A multivariate robust control chart for individual observations.
\newblock {\em Journal of Quality Technology}, 41(3):259--271.

\bibitem[Chicken et~al., 2009]{chicken2009statistical}
Chicken, E., Pignatiello~Jr, J.~J., and Simpson, J.~R. (2009).
\newblock Statistical process monitoring of nonlinear profiles using wavelets.
\newblock {\em Journal of Quality Technology}, 41(2):198--212.

\bibitem[Chou et~al., 2014]{chou2014monitoring}
Chou, S.-H., Chang, S.~I., and Tsai, T.-R. (2014).
\newblock On monitoring of multiple non-linear profiles.
\newblock {\em International Journal of Production Research}, 52(11):3209--3224.

\bibitem[Colosimo et~al., 2022]{colosimo2022complex}
Colosimo, B.~M., Grasso, M., Garghetti, F., and Rossi, B. (2022).
\newblock Complex geometries in additive manufacturing: A new solution for lattice structure modeling and monitoring.
\newblock {\em Journal of Quality Technology}, 54(4):392--414.

\bibitem[Eslami et~al., 2023]{eslami2023statistical}
Eslami, D., Izadbakhsh, H., Ahmadi, O., and Zarinbal, M. (2023).
\newblock Statistical monitoring of image data using multi-channel functional principal component analysis.
\newblock {\em Communications in Statistics-Theory and Methods}, 52(12):4165--4182.

\bibitem[Ferraty and Vieu, 2006]{ferraty2006nonparametric}
Ferraty, F. and Vieu, P. (2006).
\newblock {\em Nonparametric functional data analysis: theory and practice}.
\newblock Springer Science \& Business Media.

\bibitem[Fisher, 1992]{fisher1992statistical}
Fisher, R.~A. (1992).
\newblock Statistical methods for research workers.
\newblock In {\em Breakthroughs in statistics}, pages 66--70. Springer.

\bibitem[Grasso et~al., 2016]{grasso2016using}
Grasso, M., Menafoglio, A., Colosimo, B.~M., and Secchi, P. (2016).
\newblock Using curve-registration information for profile monitoring.
\newblock {\em Journal of Quality Technology}, 48(2):99.

\bibitem[Happ and Greven, 2018]{happ2018multivariate}
Happ, C. and Greven, S. (2018).
\newblock Multivariate functional principal component analysis for data observed on different (dimensional) domains.
\newblock {\em Journal of the American Statistical Association}, 113(522):649--659.

\bibitem[Horv{\'a}th and Kokoszka, 2012]{horvath2012inference}
Horv{\'a}th, L. and Kokoszka, P. (2012).
\newblock {\em Inference for functional data with applications}.
\newblock Springer Science \& Business Media.

\bibitem[Hubert et~al., 2005]{hubert2005robpca}
Hubert, M., Rousseeuw, P.~J., and Vanden~Branden, K. (2005).
\newblock {ROBPCA}: a new approach to robust principal component analysis.
\newblock {\em Technometrics}, 47(1):64--79.

\bibitem[Jalilibal et~al., 2025]{jalilibal2025novel}
Jalilibal, Z., Amiri, A., and Ahmadi, O. (2025).
\newblock Novel control charts for joint monitoring of the mean and variability of multichannel profiles.
\newblock {\em Quality and Reliability Engineering International}.

\bibitem[Jeong et~al., 2006]{jeong2006wavelet}
Jeong, M.~K., Lu, J.-C., and Wang, N. (2006).
\newblock Wavelet-based {SPC} procedure for complicated functional data.
\newblock {\em International Journal of Production Research}, 44(4):729--744.

\bibitem[Jeong et~al., 2018]{jeong2018statistical}
Jeong, Y.-S., Jeong, M.~K., Lu, J.-C., Yuan, M., and Jin, J. (2018).
\newblock Statistical process control procedures for functional data with systematic local variations.
\newblock {\em IISE Transactions}, 50(5):448--462.

\bibitem[Kokoszka and Reimherr, 2017]{kokoszka2017introduction}
Kokoszka, P. and Reimherr, M. (2017).
\newblock {\em Introduction to functional data analysis}.
\newblock Chapman and Hall/CRC.

\bibitem[Kruger and Xie, 2012]{kruger2012statistical}
Kruger, U. and Xie, L. (2012).
\newblock {\em Statistical monitoring of complex multivariate processes: with applications in industrial process control}.
\newblock John Wiley \& Sons.

\bibitem[Liu et~al., 2024]{liu2024comprehensive}
Liu, P., Xu, H., and Zhang, C. (2024).
\newblock A comprehensive survey of recent research on profile data analysis.
\newblock {\em Journal of Quality Technology}, 56(5):428--454.

\bibitem[Liu et~al., 2023]{liu2023covariate}
Liu, X., Du, J., and Ye, Z.-S. (2023).
\newblock A covariate-regulated sparse subspace learning model and its application to process monitoring and fault isolation.
\newblock {\em Technometrics}, 65(2):269--280.

\bibitem[Mahmoud and Woodall, 2004]{mahmoud2004phase}
Mahmoud, M.~A. and Woodall, W.~H. (2004).
\newblock Phase {I} analysis of linear profiles with calibration applications.
\newblock {\em Technometrics}, 46(4):380--391.

\bibitem[Maleki et~al., 2018]{maleki2018overview}
Maleki, M.~R., Amiri, A., and Castagliola, P. (2018).
\newblock An overview on recent profile monitoring papers (2008--2018) based on conceptual classification scheme.
\newblock {\em Computers \& Industrial Engineering}, 126:705--728.

\bibitem[Menafoglio et~al., 2018]{menafoglio2018profile}
Menafoglio, A., Grasso, M., Secchi, P., and Colosimo, B.~M. (2018).
\newblock Profile monitoring of probability density functions via simplicial functional pca with application to image data.
\newblock {\em Technometrics}, 60(4):497--510.

\bibitem[Montgomery, 2007]{montgomery2007introduction}
Montgomery, D.~C. (2007).
\newblock {\em Introduction to statistical quality control}.
\newblock John Wiley \& Sons.

\bibitem[Noorossana et~al., 2011]{noorossana2011statistical}
Noorossana, R., Saghaei, A., and Amiri, A. (2011).
\newblock {\em Statistical analysis of profile monitoring}.
\newblock John Wiley \& Sons.

\bibitem[Paynabar and Jin, 2011]{paynabar2011characterization}
Paynabar, K. and Jin, J. (2011).
\newblock Characterization of non-linear profiles variations using mixed-effect models and wavelets.
\newblock {\em IIE Transactions}, 43(4):275--290.

\bibitem[Paynabar et~al., 2016]{paynabar2016change}
Paynabar, K., Zou, C., and Qiu, P. (2016).
\newblock A change-point approach for {P}hase-{I} analysis in multivariate profile monitoring and diagnosis.
\newblock {\em Technometrics}, 58(2):191--204.

\bibitem[Qiu, 2013]{qiu2013introduction}
Qiu, P. (2013).
\newblock {\em Introduction to statistical process control}.
\newblock CRC press.

\bibitem[Qiu and Zou, 2010]{qiu2010control}
Qiu, P. and Zou, C. (2010).
\newblock Control chart for monitoring nonparametric profiles with arbitrary design.
\newblock {\em Statistica Sinica}, pages 1655--1682.

\bibitem[Qiu et~al., 2010]{qiu2010nonparametric}
Qiu, P., Zou, C., and Wang, Z. (2010).
\newblock Nonparametric profile monitoring by mixed effects modeling.
\newblock {\em Technometrics}, 52(3):265--277.

\bibitem[Rajabi et~al., 2017]{rajabi2017phase}
Rajabi, M., Faridrohani, M.~R., and Chenouri, S. (2017).
\newblock Phase {I} monitoring with nonparametric mixed-effect models.
\newblock {\em Quality and Reliability Engineering International}, 33(8):1929--1941.

\bibitem[Ramaker et~al., 2004]{ramaker2004effect}
Ramaker, H.-J., van Sprang, E.~N., Westerhuis, J.~A., and Smilde, A.~K. (2004).
\newblock The effect of the size of the training set and number of principal components on the false alarm rate in statistical process monitoring.
\newblock {\em Chemometrics and Intelligent Laboratory Systems}, 73(2):181--187.

\bibitem[Ramsay, 2005]{ramsay2005functional}
Ramsay, J.~O. (2005).
\newblock {\em Functional data analysis}.
\newblock Wiley Online Library.

\bibitem[Ren et~al., 2019]{ren2019phase}
Ren, H., Chen, N., and Wang, Z. (2019).
\newblock Phase-{II} monitoring in multichannel profile observations.
\newblock {\em Journal of Quality Technology}, 51(4):338--352.

\bibitem[Scheipl et~al., 2015]{scheipl2015functional}
Scheipl, F., Staicu, A.-M., and Greven, S. (2015).
\newblock Functional additive mixed models.
\newblock {\em Journal of Computational and Graphical Statistics}, 24(2):477--501.

\bibitem[Scimone et~al., 2022]{scimone2022statistical}
Scimone, R., Taormina, T., Colosimo, B.~M., Grasso, M., Menafoglio, A., and Secchi, P. (2022).
\newblock Statistical modeling and monitoring of geometrical deviations in complex shapes with application to additive manufacturing.
\newblock {\em Technometrics}, 64(4):437--456.

\bibitem[Shamp et~al., 2021]{shamp2021computationally}
Shamp, W., Varbanov, R., Chicken, E., Linero, A., and Yang, Y. (2021).
\newblock Computationally efficient bayesian sequential function monitoring.
\newblock {\em Journal of Quality Technology}, 54(1):1--19.

\bibitem[Tippett et~al., 1931]{tippett1931methods}
Tippett, L. H.~C. et~al. (1931).
\newblock The methods of statistics.
\newblock {\em The Methods of Statistics.}

\bibitem[Vargas, 2003]{vargas2003robust}
Vargas, N. J.~A. (2003).
\newblock Robust estimation in multivariate control charts for individual observations.
\newblock {\em Journal of Quality Technology}, 35(4):367--376.

\bibitem[Wang and Tsung, 2005]{wang2005using}
Wang, K. and Tsung, F. (2005).
\newblock Using profile monitoring techniques for a data-rich environment with huge sample size.
\newblock {\em Quality and Reliability Engineering International}, 21(7):677--688.

\bibitem[Wang and Tsung, 2021]{wang2021sparse}
Wang, K. and Tsung, F. (2021).
\newblock Sparse and robust multivariate functional principal component analysis for passenger flow pattern discovery in metro systems.
\newblock {\em IEEE Transactions on Intelligent Transportation Systems}, 23(7):8367--8379.

\bibitem[Wang et~al., 2022]{wang2022outlier}
Wang, T., Wang, Y., and Zang, Q. (2022).
\newblock Outlier detection in non-parametric profile monitoring.
\newblock {\em Statistics}, 56(4):805--822.

\bibitem[Wang et~al., 2018]{wang2018thresholded}
Wang, Y., Mei, Y., and Paynabar, K. (2018).
\newblock Thresholded multivariate principal component analysis for phase {I} multichannel profile monitoring.
\newblock {\em Technometrics}, 60(3):360--372.

\bibitem[Williams et~al., 2007]{williams2007statistical}
Williams, J.~D., Woodall, W.~H., and Birch, J.~B. (2007).
\newblock Statistical monitoring of nonlinear product and process quality profiles.
\newblock {\em Quality and Reliability Engineering International}, 23(8):925--941.

\bibitem[Wittenberg et~al., 2025]{wittenberg2025covariate}
Wittenberg, P., Neumann, L., Mendler, A., and Gertheiss, J. (2025).
\newblock Covariate-adjusted functional data analysis for structural health monitoring.
\newblock {\em Data-Centric Engineering}, 6:e27.

\bibitem[Woodall et~al., 2004]{woodall2004using}
Woodall, W.~H., Spitzner, D.~J., Montgomery, D.~C., and Gupta, S. (2004).
\newblock Using control charts to monitor process and product quality profiles.
\newblock {\em Journal of Quality Technology}, 36(3):309.

\bibitem[Wu et~al., 2022]{wu2022monitoring}
Wu, H., Zhang, C., and Li, Y.-F. (2022).
\newblock Monitoring heterogeneous multivariate profiles based on heterogeneous graphical model.
\newblock {\em Technometrics}, 64(2):210--223.

\bibitem[Yan et~al., 2018]{yan2018real}
Yan, H., Paynabar, K., and Shi, J. (2018).
\newblock Real-time monitoring of high-dimensional functional data streams via spatio-temporal smooth sparse decomposition.
\newblock {\em Technometrics}, 60(2):181--197.

\bibitem[Yao et~al., 2024]{yao2024adaptive}
Yao, J., Balasubramaniam, B., Li, B., Kreiger, E.~L., and Wang, C. (2024).
\newblock Adaptive sampling and monitoring of partially observed images.
\newblock {\em Journal of Quality Technology}, 56(2):157--173.

\bibitem[Yao et~al., 2023]{yao2023adaptive}
Yao, J., Xian, X., and Wang, C. (2023).
\newblock Adaptive sampling for monitoring multi-profile data with within-and-between profile correlation.
\newblock {\em Technometrics}, 65(3):375--387.

\bibitem[Yu et~al., 2012]{yu2012outlier}
Yu, G., Zou, C., and Wang, Z. (2012).
\newblock Outlier detection in functional observations with applications to profile monitoring.
\newblock {\em Technometrics}, 54(3):308--318.

\bibitem[Zang et~al., 2016]{zang2016unaligned}
Zang, Y., Wang, K., and Jin, R. (2016).
\newblock Unaligned profile monitoring using penalized methods.
\newblock {\em Quality and Reliability Engineering International}, 32(8):2761--2776.

\bibitem[Zhang et~al., 2018a]{zhang2018multiple}
Zhang, C., Yan, H., Lee, S., and Shi, J. (2018a).
\newblock Multiple profiles sensor-based monitoring and anomaly detection.
\newblock {\em Journal of Quality Technology}, 50(4):344--362.

\bibitem[Zhang et~al., 2018b]{zhang2018weakly}
Zhang, C., Yan, H., Lee, S., and Shi, J. (2018b).
\newblock Weakly correlated profile monitoring based on sparse multi-channel functional principal component analysis.
\newblock {\em IISE Transactions}, 50(10):878--891.

\bibitem[Zhang et~al., 2015]{zhang2015phase}
Zhang, J., Ren, H., Yao, R., Zou, C., and Wang, Z. (2015).
\newblock Phase i analysis of multivariate profiles based on regression adjustment.
\newblock {\em Computers \& Industrial Engineering}, 85:132--144.

\bibitem[Zhou and Qiu, 2022]{zhou2022phase}
Zhou, Q. and Qiu, P. (2022).
\newblock Phase i monitoring of serially correlated nonparametric profiles by mixed-effects modeling.
\newblock {\em Quality and Reliability Engineering International}, 38(1):134--152.

\bibitem[Zhou et~al., 2006]{zhou2006spc}
Zhou, S., Sun, B., and Shi, J. (2006).
\newblock An {SPC} monitoring system for cycle-based waveform signals using haar transform.
\newblock {\em IEEE Transactions on Automation Science and Engineering}, 3(1):60--72.

\bibitem[Zou et~al., 2012]{zou2012lasso}
Zou, C., Ning, X., and Tsung, F. (2012).
\newblock {LASSO}-based multivariate linear profile monitoring.
\newblock {\em Annals of operations research}, 192(1):3--19.

\bibitem[Zou et~al., 2007]{zou2007monitoring}
Zou, C., Tsung, F., and Wang, Z. (2007).
\newblock Monitoring general linear profiles using multivariate exponentially weighted moving average schemes.
\newblock {\em Technometrics}, 49(4):395--408.

\bibitem[Zou et~al., 2008]{zou2008monitoring}
Zou, C., Tsung, F., and Wang, Z. (2008).
\newblock Monitoring profiles based on nonparametric regression methods.
\newblock {\em Technometrics}, 50(4):512--526.

\bibitem[Zou et~al., 2006]{zou2006control}
Zou, C., Zhang, Y., and Wang, Z. (2006).
\newblock A control chart based on a change-point model for monitoring linear profiles.
\newblock {\em IIE Transactions}, 38(12):1093--1103.

\end{thebibliography}
}

\end{document}